\documentclass[10pt, lettersize,journal]{IEEEtran}
\usepackage{amsmath,amsfonts}
\usepackage{algorithmic}
\usepackage{algorithm}
\usepackage{array}
\usepackage{textcomp}
\usepackage{stfloats}
\usepackage{url}
\usepackage{verbatim}
\usepackage{graphicx}
\usepackage{cite}
\usepackage{booktabs}
\usepackage{threeparttable}
\usepackage{multicol, multirow}
\usepackage{bm}
\usepackage{subfigure}

\def\L{{\cal L}}

\DeclareMathOperator*{\argmin}{arg\,min}
\usepackage{enumitem}

\def\D{{\cal D}}

\def\M{{\cal M}}

\begin{document}

\title{Efficient Rehearsal for Continual Learning in ASR \\ via Singular Value Tuning}

\author{Steven Vander Eeckt, Hugo Van hamme \IEEEmembership{Senior, IEEE}
\thanks{S. Vander Eeckt and H. Van hamme are with Department Electrical Engineering ESAT-PSI, KU Leuven, B-3001 Leuven, Belgium}
}

%\author{IEEE Publication Technology,~\IEEEmembership{Staff,~IEEE,}
        % <-this % stops a space
%\thanks{This paper was produced by the IEEE Publication Technology Group. They are in Piscataway, NJ.}% <-this % stops a space
%\thanks{Manuscript received April 19, 2021; revised August 16, 2021.}}



\maketitle

\begingroup
\renewcommand\thefootnote{}\footnotetext{
© 2026 IEEE. Personal use of this material is permitted. Permission from IEEE must be obtained for all other uses, in any current or future media, including reprinting/republishing this material for advertising or promotional purposes, creating new collective works, for resale or redistribution to servers or lists, or reuse of any copyrighted component of this work in other works.
}
\endgroup

\begin{abstract}
Continual Learning (CL) in Automatic Speech Recognition (ASR) suffers from catastrophic forgetting when adapting to new tasks, domains, or speakers. A common strategy to mitigate this is to store a subset of past data in memory for rehearsal. However, rehearsal-based methods face key limitations: storing data is often costly, infeasible with pre-trained models, or restricted by privacy regulations. Running existing rehearsal-based methods with smaller memory sizes to alleviate these issues usually leads to degraded performance.

We propose a rehearsal-based CL method that remains effective even with minimal memory. It operates in two stages: first, fine-tuning on the new task; second, applying Singular Value Decomposition (SVD) to the changes in linear layers and, in a parameter-efficient manner, retraining only gating vectors on the singular values, which control to extent to which updates from the first stage are accepted, using rehearsal. We extensively test and analyze our method on two monolingual and two multilingual benchmarks. Our method reduces forgetting and outperforms state-of-the-art CL approaches for ASR, even when limited to a single utterance per previous task.
\end{abstract}


\section{Introduction}
Automatic Speech Recognition (ASR) systems are increasingly deployed in real-world, dynamic environments where they must continuously adapt to new speakers, domains, accents or conditions. In such settings, continual learning (CL) becomes essential as it allows models to incrementally learn new tasks without access to previously seen data. However, a major challenge in CL is \textit{catastrophic forgetting}---the tendency of neural networks to lose performance on previously learned tasks when updated with new data~\cite{catastrophicforgetting}.

A prominent strategy to mitigate forgetting is \textit{rehearsal}, where a small subset of past data is retained and interleaved with new task data during training~\cite{icarl, gem}. \textit{Rehearsal-based methods} have proven effective in ASR~\cite{eeckt2021continual, lifelongasr}, but come with  limitations: storing real user data may be infeasible due to privacy or regulatory constraints, and managing large memories can be computationally prohibitive. These issues motivate the development of \textit{memory-efficient} rehearsal methods that can operate with very limited memory.

In this work, we introduce a rehearsal-based continual learning method for ASR that is both effective and highly memory-efficient. Our approach involves two stages. First, we fine-tune the model on the new task. Then, we perform a structured update using Singular Value Decomposition (SVD) of the linear layer weight changes. By introducing a learnable gating vector on the singular values, we selectively retain only those updates that balance learning the new task with preserving performance on previous ones. Importantly, only a small number of parameters are updated during this stage, making our method well-suited to low-memory settings.

We extensively evaluate our method across four experiments: two monolingual setups involving microphone and accent shifts, one multilingual setup with task-specific model components, and one multilingual setup based on a Whisper-style \cite{whisper} foundation model. The results show that our method is highly effective even when using an extremely small memory—down to a single utterance per task—and consistently outperforms state-of-the-art baselines, including both non-rehearsal methods and rehearsal-based approaches that rely on significantly larger memory budgets. Our method achieves a strong balance between retaining past knowledge and adapting to new tasks. In addition, we conduct an in-depth analysis of the learned gating vectors, which reveals that the model tends to either fully suppress or fully accept individual rank-one updates. Finally, an ablation study confirms the importance of our method's key components toward its overall performance. 

\section{Continual Learning for ASR}
\subsection{Model and Notation}
The ASR model used in this work is an encoder-decoder architecture trained in a hybrid fashion~\cite{hybrid_ctctransformer}. It takes as input an utterance \( \bm{X} \in \mathbb{R}^{l \times d_i} \), where \( l \) is the number of acoustic frames and \( d_i \) is the dimensionality of each frame. The output is a sequence \( \bm{\hat{y}} \) of \( \tilde{w} \) predicted word pieces.

The model parameters are denoted by \( \bm{\theta} \in \mathbb{R}^N \). The training objective combines two losses: a CTC loss with weight \( c \), and a decoder-based cross-entropy loss with weight \( 1 - c \). The model is trained on paired data \( (\bm{X}, \bm{y}) \), where \( \bm{y} \) is the ground truth sequence of \( w \) word pieces corresponding to input \( \bm{X} \). Eq. \ref{eq:loss} denotes the loss of the model:
\begin{equation}
    \L_\text{ce}(\bm{X}, \bm{y}; \bm{\theta}) = c\L_\text{ctc}(\bm{X}, \bm{y}; \bm{\theta}) + (1 - c)\L_\text{dec}(\bm{X}, \bm{y}; \bm{\theta})
    \label{eq:loss}
\end{equation}
With $\L_\text{ctc}(\cdot)$ and $\L_\text{dec}(\cdot)$ the CTC loss and decoder-based cross-entropy loss, respectively.

We denote the output distributions (after applying the softmax) of the two branches as follows: \( f^{\text{ctc}}(\bm{X}; \bm{\theta}) \in \mathbb{R}^{l \times C} \) is the CTC output; \( f^{\text{dec}}(\bm{X}; \bm{\theta}) \in \mathbb{R}^{w \times C} \) is the decoder output, computed during training and conditioned on previous ground-truth tokens, with $C$ the output size and size of the vocabulary.

\subsection{Problem Formulation}

We consider a continual learning (CL) setting for ASR where a model must learn a sequence of $T$ tasks, each becoming available one at a time. Tasks may vary in language, speaker, accent, recording environment, etc. For each task $t \in \{1, \dots, T\}$, we are given a training set $\D^\text{train}_t$ and a validation set $\D^\text{val}_t$, both consisting of paired utterances $(\bm{X}, \bm{y})$. Once task $t$ becomes available, access to the training and validation sets of the previous task $t{-}1$ is lost—that is, $\D^\text{train}_{t-1}$ and $\D^\text{val}_{t-1}$ are no longer accessible.

Given a model with parameters $\bm{\theta}^{t-1} \in \mathbb{R}^N$ performing well on tasks $1, \dots, t{-}1$, the goal of CL is to obtain updated parameters $\bm{\theta}^t$ (using $\D^{\text{train}}_t$ and $\D^{\text{val}}_t$) that satisfy two criteria~\cite{biesialska-etal-2020-continual}:

\begin{enumerate}[label=(\roman*)]
    \item \textit{Retain previously learned knowledge}: Maintain performance on all previous tasks. Ideally, for each $j \in \{1, \dots, t{-}1\}$,
    \begin{equation}
        \sum_{(\bm{X}, \bm{y}) \in \D^\text{train}_j} \mathcal{L}_\text{ce}(\bm{X}, \bm{y}; \bm{\theta}^t) \leq \sum_{(\bm{X}, \bm{y}) \in \D^\text{train}_j} \mathcal{L}_\text{ce}(\bm{X}, \bm{y}; \bm{\theta}^{t-1}),
        \label{eq:knowledge_retention}
    \end{equation}
    \item \textit{Adapt effectively to new data}: Perform well on the new task $t$. That is, as a proxy:
    \begin{equation}
        \bm{\theta}^t = \argmin_{\bm{\theta}} \sum_{(\bm{X}, \bm{y}) \in \D^\text{train}_t} \mathcal{L}_\text{ce}(\bm{X}, \bm{y}; \bm{\theta}).
        \label{eq:forward_transfer}
    \end{equation}
\end{enumerate}

To meet these objectives, CL methods typically modify the optimization procedure, adjust the architecture, or incorporate a small memory buffer $\mathcal{M}$ containing $|\mathcal{M}|$ samples from previous tasks. \text{Rehearsal-based methods} use $\mathcal{M}$ to replay past samples during training on task $t$, mitigating forgetting by approximating access to $\D^\text{train}_{1}, \dots, \D^\text{train}_{t-1}$. However, storing a large memory may be impractical—due to privacy constraints, storage limitations, or restrictions when using pretrained models. As such, designing methods that remain effective even with a highly limited memory is crucial, especially as naively fine-tuning $\bm{\theta}^{t-1}$ on task $t$ without such strategies generally results in catastrophic forgetting.

\subsection{Related Work}

\noindent \textbf{Continual Learning.} Following \cite{defy}, CL methods can be categorized into three groups. \textit{Regularization-based methods}, e.g. \cite{ewc, mas, imm, lwf}, alleviate forgetting through auxiliary loss terms that penalize changes to important parameters. \textit{Architectural-based methods}, e.g., \cite{hat, progresscompress}, modify the model architecture to allocate task-specific capacity. Finally, \textit{Rehearsal-based methods} store a small memory of training examples from past tasks and interleave them with current task data to mitigate forgetting. Despite their simplicity, rehearsal-based strategies remain among the most effective approaches in continual learning \cite{icarl, gem, agem, er, dark_er}. \cite{np_hard} shows that optimal continual learning is NP-hard and would require perfect memorization of past tasks, making approximate recall strategies such as experience replay particularly promising \cite{comprehensive_review}. Their theoretical analysis supports the empirical success and widespread use of rehearsal-based methods \cite{rehearsal_review} over regularization-based ones. The effectiveness of rehearsal-based methods has also been attributed to their ability to remain within low-loss regions of previously learned tasks~\cite{rehearsal_review}.

\noindent \textbf{Continual Learning for ASR.} In recent years, CL has gained increasing attention within the ASR community. \cite{lifelongasr, eeckt2021continual} show that {rehearsal-based methods} significantly outperform {regularization-based methods}, highlighting the importance of a memory of past samples. To omit this memory requirement, \cite{eeckt_adapters} considers task-specific adapters \cite{adapters}, while \cite{sustek22_interspeech} combines expert models, and \cite{disentangled} introduces a disentangled architecture with task-specific parameters—examples of {architectural-based strategies}. {Regularization-based methods} for ASR were proposed by \cite{weight_averaging}, which introduces weight averaging, illustrating its efficacy for CL in ASR; \cite{updating_only}, which proposes updating only the encoder for CL; and \cite{wang23d_interspeech}, which suggests to update only a subset of encoders during each epoch. \cite{kwok24_interspeech, xu24h_interspeech, kwok} focus on CL for multilingual Whisper-based \cite{whisper} ASR, with the latter two adopting low-rank adaptation (LoRA) \cite{lora} within architectural-based methods. \cite{other_ucl, ogem, vandereeckt_interspeech2023, vandereeckt24_interspeech} explore online CL for ASR, where the model is exposed to a stream of non-i.i.d. data rather than a sequence of tasks with known boundaries. Online CL is not considered in this paper.

\noindent \textbf{SVD-Based Learning.} SVD-based learning has been explored within the context of parameter-efficient fine-tuning of pretrained models. \cite{adalora, increlora} consider LoRA \cite{lora} decomposed via SVD. \cite{kopiczko2024vera} further reduces the number of trainable parameters by fine-tuning only scale vectors. \cite{curlora} considers the adaptation presented in its CUR decomposition \cite{mahoney2009cur}, training only the matrix $\bm{U}$. \cite{svf_neurips, svft_neurips2} consider singular-value fine-tuning. \cite{corda} performs SVD on the product of pretrained weights and task-specific covariance matrices, orienting the decomposition toward task context and using it to initialize LoRA. Note that in these works, the focus is on adapting a pretrained model to a new task, not on preventing forgetting as in our work.

\section{Singular Value-based Rehearsal}

We propose a memory-efficient rehearsal-based continual learning method (which we name Singular Value-based Rehearsal [SVR]) for ASR, consisting of two stages. Given a model trained on the previous $t-1$ tasks with parameters $\bm{\theta}^{t-1}$, the objective is to train $\bm{\theta}$ on task $t$, using $\bm{\theta}^{t-1}$, $\D^\text{train}_t$ and $\M$, in order to obtain $\bm{\theta}^t \gets \bm{\theta}$ which works well on all seen tasks $1, \dots, t$.

\subsection{Stage 1: Fine-Tuning to learn the new task $t$}
\label{subsec:methodst1}
 We first perform conventional fine-tuning on the new task $t$ using $\D_t^\text{train}$, starting from model parameters $\bm{\theta}^{t-1}$. This results in an updated set of parameters $\bm{\tilde{\theta}}^t$. Let $\Delta\bm{\tilde{\theta}}_t = \bm{\tilde{\theta}}^t - \bm{\theta}^{t-1}$ denote the parameter changes induced by this adaptation.

 Although effective for the current task, $\Delta\bm{\tilde{\theta}}_t$ often degrades performance on previously learned tasks, resulting in catastrophic forgetting. To address this, we introduce a second stage that filters out components of $\Delta\bm{\tilde{\theta}}_t$ that are detrimental to tasks $1, \dots, t-1$, yielding a model that maintains performance on both past and current tasks.

Note that the memory $\mathcal{M}$ is not used during this stage.

\subsection{Stage 2a: Preparing Singular Value-Based Rehearsal}
\label{subsec:methodst2}
Consider a linear layer in the ASR model with parameters $\bm{\theta}$, represented by a weight matrix $\bm{W} \in \mathbb{R}^{d_o \times d_i}$ and a bias vector $\bm{b} \in \mathbb{R}^{d_o}$. Given an input $\bm{h} \in \mathbb{R}^{d_i}$, the output of the layer is computed as $(\bm{W}\bm{h} + \bm{b}) \in \mathbb{R}^{d_o}$. We hypothesize that the weight matrix $\bm{W}$ plays the primary role in both learning new tasks and forgetting old ones, whereas the bias $\bm{b}$ contributes only marginally. Therefore, our focus is on achieving an effective balance between retaining knowledge of previous tasks and adapting to new ones by carefully managing updates to $\bm{W}$.

Let $\bm{\tilde{W}}_t$ denote the linear layer of the model with parameters $\bm{\tilde{\theta}}^t$ after fine-tuning on task $t$ (Stage 1 from Sec. \ref{subsec:methodst1}), and let $\bm{W}_{t-1}$ denote the corresponding layer in the model with parameters $\bm{\theta}_{t-1}$ prior to adaptation. We define the weight update as $\Delta \bm{W}_t = \bm{\tilde{W}}_t - \bm{W}_{t-1}$. Writing this update $\Delta \bm{W}_t$, which is applied to $\bm{W}_{t-1}$ to adapt to the new task, in terms of its Singular Value Decomposition (SVD), we have:
\begin{equation}
\Delta \bm{W}_t = \bm{U} \bm{\Sigma} \bm{V}^T = \sum_{i=1}^k s_i \bm{u}_i \bm{v}_i^T
\end{equation}
where $k \leq \min(d_o, d_i)$. Here, $\bm{U} \in \mathbb{R}^{d_o \times k}$ and $\bm{V} \in \mathbb{R}^{d_i \times k}$ contain the left and right singular vectors, respectively, with $\bm{u}_i \in \mathbb{R}^{d_o}$ and $\bm{v}_i \in \mathbb{R}^{d_i}$ denoting the $i$-th columns of $\bm{U}$ and $\bm{V}$. The diagonal matrix $\bm{\Sigma} \in \mathbb{R}^{k \times k}$ holds the nonnegative singular values $\bm{s}_i$ on its diagonal. Each rank-one component $s_i \bm{u}_i \bm{v}_i^T$ represents a principal direction of change induced by the update $\Delta \bm{W}_t$.

The update $\Delta \bm{W}_t$ can thus be expressed as a sum of $k$ rank-one components: $s_i \bm{u}_i \bm{v}_i^T$ for $i = 1, \dots, k$. Importantly, not all of these components contribute equally to learning the new task or to forgetting previously learned tasks. Ideally, we would retain those updates that significantly improve performance on the new task while causing minimal interference with earlier tasks, and discard those that contribute little to learning but induce substantial forgetting. For updates that are beneficial for the new task but also harmful to prior tasks, a compromise may be achieved through averaging \cite{weight_averaging}. To implement such selective control over the influence of each component, we introduce a weighting vector $\bm{\alpha} \in \mathbb{R}^k$ and define a modified update $\Delta \bm{\hat{W}}_t$ as:
\begin{equation}
    \Delta \bm{\hat{W}}_t = \bm{U}\text{diag}(\sigma(\bm{\alpha}) \odot \bm{s})\bm{V}^T = \sum_{i=1}^k \sigma(\alpha_i)s_i \bm{u}_i \bm{v}_i^T
\end{equation}
where $\sigma(\cdot)$ is the sigmoid function, $\bm{s} \in \mathbb{R}^k$ is the diagonal of $\bm{\Sigma}$, $\text{diag}(\bm{x})$ turns $\bm{x} \in \mathbb{R}^d$ into a $d\times d$ diagonal matrix, and $\odot$ denotes element-wise multiplication. The updated linear layer in the current model $\bm{\theta}$ is computed as:
\begin{equation}
\bm{W} \gets \bm{W}_{t-1} + \Delta \bm{\hat{W}}_t = \bm{W}_{t-1} + \bm{U} \text{diag}\left( \sigma(\bm{\alpha}) \odot \bm{s} \right) \bm{V}^T
\end{equation}
Here, $\bm{\alpha} \in \mathbb{R}^k$ is a trainable vector, and each $\sigma(\alpha_i) \in [0, 1]$ controls the contribution of the $i$-th rank-one update. During training, only $\bm{\alpha}$ is updated, while $\bm{U}$, $\bm{\Sigma}$ (thus $\bm{s}$), $\bm{V}$, and $\bm{W}_{t-1}$ remain fixed. 

To achieve a balance between retaining performance on previous tasks and adapting to the new task, $\bm{\alpha}$ is optimized using rehearsal: it is trained jointly on the memory of old samples and the new task data. Through this \textit{Stage 2 training}, the model learns to selectively scale the rank-one updates proposed by Stage 1. If a particular update negatively impacts prior tasks, the model can suppress it by driving $\sigma(\alpha_i) \approx 0$. Conversely, if an update is beneficial and causes minimal forgetting, it may be preserved with $\sigma(\alpha_i) \approx 1.0$. In cases where an update benefits the new task but harms the old ones, the model may seek a trade-off by setting $\sigma(\alpha_i) \approx 0.5$.

The proposed SVD-based rehearsal strategy can be applied to all linear layers in the model, which collectively account for the vast majority of the model’s parameters. All other parameters, including those in bias vectors, and convolutional and embedding layers, are kept frozen throughout this stage. However, motivated by the effectiveness of weight averaging \cite{weight_averaging}, we initialize these other model parameters (to which our SVD-based rehearsal is \textit{not} applied) as the average of their values in $\bm{\theta}^{t-1}$ and $\bm{\tilde{\theta}}^t$. For instance, if $\bm{p}$ is such a model parameter in the current model, and $\bm{\tilde{p}}_t$ and $\bm{p}_{t-1}$ are the corresponding model parameters from $\bm{\tilde{\theta}}^t$ and $\bm{\theta}^{t-1}$, respectively, we set $\bm{p}$ as follows:
\begin{equation}
    \bm{p}\gets \frac{\bm{\tilde{p}}_t+ \bm{p}_{t-1}}{2}
\end{equation}
$\bm{p}$ then remains frozen. As the number of trainable parameters is thus limited to the size of all $\bm\alpha$'s, our approach offers substantial efficiency: for each linear layer, only $k \leq \min(d_o, d_i)$ parameters in $\sigma(\bm{\alpha})$ are trained, as opposed to the full $d_o \times d_i$ weight matrix. We hypothesize that this reduction in parameter count allows Stage 2 to be trained effectively using a minimal memory buffer, minimizing the risk of overfitting on the memory while mitigating forgetting of old tasks.

\subsection{Stage 2b: Singular Value-Based Rehearsal}
\label{subsec:methodst2b}
To train $\sigma(\bm{\alpha})$ for each linear layer to which our SVD-based rehearsal is applied—jointly on the memory of old tasks and the new task—we minimize the following loss:

\begin{equation}
    \begin{aligned}
    \L_\text{st2}(\bm{X}, \bm{y}, &\bm{X}_M, \bm{y}_M; \bm{\theta}) =\L_\text{ce}(\bm{X},\bm{y}; \bm{\theta})  \\ 
    & + \frac{t-1}{2} \left( \L_\text{ce}(\bm{X}_M, \bm{y}_M; \bm{\theta}) + \L_\text{kd}(\bm{X}_M; \bm{\theta}) \right)
    \end{aligned}
    \label{eq:loss_st2}
\end{equation}
with $(\bm{X}, \bm{y}) \sim D^\text{train}_t$, $(\bm{X}_M, \bm{y}_M) \sim \mathcal{M}$, while $\L_\text{kd}$ is the knowledge distillation \cite{knowledge_distillation, eeckt2021continual} loss given by Eq. \ref{eq:loss_kd}:
\begin{equation}
    \begin{aligned}
    \L_\text{kd} & (\bm{X}; \bm{\theta}) = c \sum_{i=1}^l \sum_{j=1}^C {f_{i,j}^\text{ctc}(\bm{X}; \bm{\theta}^{t-1})} \log {f_{i,j}^\text{ctc}(\bm{X}; \bm{\theta})} \\
     & + (1 - c) \sum_{i=1}^w \sum_{j=1}^C {f_{i,j}^\text{dec}(\bm{X}; \bm{\theta}^{t-1})} \log {f_{i,j}^\text{dec}(\bm{X};\bm{\theta})} 
    \end{aligned}
    \label{eq:loss_kd}
\end{equation}
Note that $(\bm{X}, \bm{y})$ and $(\bm{X}_M, \bm{y}_M)$ here represent mini-batches of size $b$. The loss is thus computed jointly on a mini-batch from the new task’s training set and a mini-batch from the memory buffer. For the mini-batch $(\bm{X}_M, \bm{y}_M)$ from memory, similar to \cite{dark_er}, we include both a cross-entropy loss and a knowledge distillation loss to encourage the current model to preserve the behavior of the previous model ($\bm{\theta}^{t-1}$). The memory loss is weighted by $(t-1)/2$: the division by 2 accounts for the presence of two loss components (cross-entropy and distillation), while the multiplication by $(t-1)$ reflects the fact that the memory represents all $t-1$ previous tasks. Intuitively, if one were to maintain a separate memory buffer and corresponding loss term for each prior task, this would result in $t-1$ such terms. 

At the start of training, we initialize $\bm{\alpha}$ such that $\sigma(\bm{\alpha}) \approx 0$, ensuring that initially $\bm{W} \approx \bm{W}_{t-1}$. This initialization preserves the model’s performance on the previously learned tasks $1, \dots, t-1$. It allows our method to stay within the low-loss region of old tasks (from which it does not necessarily start when $\sigma(\bm{\alpha}) > 0$) during the trajectory of training, which is generally the case for rehearsal-based methods and  partly explains their effectiveness \cite{rehearsal_review}.

\subsection{Summary}
Algorithm \ref{alg:overview} summarizes our approach. Line \ref{line:overview_st1} represents Stage 1 from Sec. \ref{subsec:methodst1}, in which we fine-tune $\bm{\theta}^{t-1}$ on $\D^\text{train}_t$, obtaining $\bm{\tilde{\theta}}^t$. %As explained in Sec. \ref{subsec:methodst1}, optionally, a lightweight CL method instead of Fine-Tuning could be used in Stage 1. 
For Stage 2, we first apply, as explained in \ref{subsec:methodst2}, SVD to and introduce $\bm{\alpha}$ for each linear layer $\bm{W}$ of $\bm{\theta}$ (Lines \ref{line:overview_st2a_start}-\ref{line:overview_st2a_end}). The remaining model parameters are averaged and frozen (Lines \ref{line:averaging_start}-\ref{line:averaging_end}). For each mini-batch of the new task's training set $\D^\text{train}_t$ and mini-batch sampled from the memory $\M$ (Line \ref{line:overview_st2b_sample}), the loss from Eq. \ref{eq:loss_st2} (Sec. \ref{subsec:methodst2b}) is computed in Line \ref{line:overview_st2b_loss}. Note that Stage 2 training for loop (Lines \ref{line:training_start}-\ref{line:training_end}) can in practice be repeated for $e_\text{st2}$ epochs. Before the model $\bm{\theta}^t$ is returned (Line \ref{line:overview_return}), the memory $\M$ is updated by adding utterances from $\D^\text{train}_t$ (Line \ref{line:overview_post}). 

\begin{algorithm}
\caption{\textbf{S}ingular \textbf{V}alue-Based \textbf{R}ehearsal (SVR)}
\begin{algorithmic}[1]
\REQUIRE Model $\bm{\theta}^{t-1}$, new task data $\D^\text{train}_t$, memory $\mathcal{M}$
\ENSURE Updated model $\bm{\theta}^t$

\STATE \textbf{\# Stage 1: Fine-tune on new task}
\STATE $\bm{\tilde{\theta}}^t \leftarrow \text{FineTune}(\bm{\theta}^{t-1}, \D^\text{train}_t)$ \label{line:overview_st1}
\STATE \textbf{\# Stage 2: SVD-based Rehearsal}
\FOR{each linear layer $\bm{W}$ in $\bm{\theta}$} \label{line:overview_st2a_start}
    \STATE {\# $\bm{\tilde{W}}_t$ and $\bm{W}_{t-1}$ are $\bm{W}$ from $\bm{\tilde{\theta}}^t$ and $\bm{\theta}^{t-1}$, resp. }
    \STATE $\Delta \bm{W}_t \leftarrow \tilde{\bm{W}}_t - \bm{W}_{t-1}$
    \STATE $\bm{U}, \bm{\Sigma}, \bm{V}^T \leftarrow \text{SVD}(\Delta \bm{W}_t)$
    %\STATE {\# Initialize $\sigma(\alpha) \approx 0$.}
    \STATE {\# Introduce trainable vector $\bm{\alpha}$, $\bm{s}$ is diagonal of $\bm{\Sigma}$}
    \STATE $\bm{W} \gets \bm{W}_{t-1} + \bm{U} \text{diag}(\sigma(\bm{\alpha}) \odot \bm{s})\bm{V}^T$
    \STATE {Freeze $\bm{W}_{t-1}$, $\bm{U}$, $\bm{\Sigma}$ (thus $\bm{s}$), $\bm{V}^T$. Train only $\bm{\alpha}$.} 
\ENDFOR \label{line:overview_st2a_end}
\STATE {\# Average all other parameters}
\FOR{each other parameter $\bm{p}$ in $\bm{\theta}$} \label{line:averaging_start}
    \STATE {\# $\tilde{\bm{p}}_t$ and $\bm{p}_{t-1}$ are parameter $\bm{p}$ from $\bm{\tilde{\theta}}^t$ and $\bm{\theta}^{t-1}$, resp.}
    \STATE $\bm{p} \gets (\bm{p}_{t-1} + \bm{\tilde{p}}_t)/2$ \label{line:averaging}
    \STATE Freeze $\bm{p}$
\ENDFOR \label{line:averaging_end}
\STATE {\# Jointly train on new task and memory}
\FOR{each mini-batch $(\bm{X}, \bm{y}) \sim \D^{\text{train}}_t$} \label{line:training_start}
    \STATE Sample $(\bm{X}_M, \bm{y}_M) \sim \M$ from memory \label{line:overview_st2b_sample}
    \STATE Compute $\L_{st2}(\bm{X}, \bm{y}, \bm{X}_M, \bm{y}_M; \bm{\theta})$ using Eq. \ref{eq:loss_st2} \label{line:overview_st2b_loss}
    \STATE Update $\bm{\theta}$ (thus all $\bm{\alpha}$'s).
\ENDFOR \label{line:training_end}
\STATE {\# Add samples from $\D^\text{train}_t$ to $\M$}
\STATE $\M \gets \text{UpdateMemory}(\M, \D^\text{train}_t)$ \label{line:overview_post}
\RETURN $\bm{\theta}^t \gets \bm{\theta}$ \label{line:overview_return}
\end{algorithmic}
\label{alg:overview}
\end{algorithm}

To populate the memory $\mathcal{M}$ with samples from the task $t$, using $\text{UpdateMemory}(\mathcal{M},\mathcal{D}^{\text{train}}_t)$ from Line~\ref{line:overview_post}, we sample uniformly from the training set $\mathcal{D}^{\text{train}}_t$.
We consider multiple memory sizes $|\mathcal{M}|$, both fixed and increasing. In the fixed setting, the memory size $|\mathcal{M}| = M$ remains constant across tasks, so each task contributes $M/t$ samples; to make room for new samples, we remove entries by uniformly sampling per task. In the increasing setting, the memory grows linearly with the number of tasks, $|\mathcal{M}| = Mt$, allowing each task to store $M$ utterances. In multilingual settings, tasks may belong to different languages. For fairness, we allocate equal total weight to each language when populating the memory:
with fixed $|\mathcal{M}|$, samples are divided equally across
languages and then across their tasks; with increasing $|\mathcal{M}| = Mt$, we subsample during training so that each language contributes equally to the regularization, regardless of its number of tasks. 

We use uniform sampling to populate the memory, as it is generally applicable,
since it does not require access to large candidate pools,
unlike more selective strategies.

\section{Experiments}
Experiments are done in ESPnet2 \cite{watanabe2018espnet}. For code and more detailed information, we refer to our Github repository \footnote{{https://github.com/StevenVdEeckt/efficient-rehearsal-for-cl-in-asr}}.

\noindent \textbf{Data.} We conduct four experiments: 
\begin{enumerate}
    \item \textit{Experiment 1}. Following \cite{weight_averaging}, we considering the Common Voice \cite{commonvoice} dataset and split the Common Voice English dataset into 5 English accents: United States (US), England (ENG), Australia (AUS), India (IND) and Scotland (SCO). The tasks are learned in this order. 
    \item \textit{Experiment 2}. Starting from a model trained on LibriSpeech-360h (LIB) \cite{librispeech}, it is adapted to four Libri-Adapt (LIB-APT) \cite{libri_adapt} tasks, with a domain shift in terms of microphone (USB [U], Matrix [M]) and accent (United States [US], Indian [IN], British [GB] English). This closely follows the set-up from \cite{vandereeckt24_interspeech}, where it served as an Online CL benchmark. 
    \item \textit{Experiment 3.} In addition to the two monolingual experiments above, we also consider one multilingual experiment, based on \cite{weight_averaging}. The tasks are five languages from the Common Voice dataset: English (US), Russian (RU), Dutch (NL), Polish (PL) and Swedish (SV). 
    \item \textit{Experiment 4}. Finally, we adapt the Open Whisper-style Speech Model (OWSM) \cite{owsm} to Dutch across two tasks. As pretrained knowledge, i.e., as old tasks, we consider Common Voice (CV) \cite{commonvoice}, both English (CV/E) and Dutch (CV/N). Performance on these tasks, which were part of the pretraining for OWSM and therefore considered {task 1}, should not deteriorate as we adapt OWSM to new tasks. These new tasks are obtained from Corpus Gesproken Nederlands (CGN) \cite{cgn}, a dataset which contains approximately 900 hours of spoken Dutch covering a wide range of speech registers not represented in CV, including spontaneous conversations, interviews, debates, lectures, and broadcast material. The CGN data are split into Dutch from the Netherlands (CG/N) and Dutch from Belgium (CG/V), using non-spontaneous speech to define the two tasks. Compared to \cite{eeckt2021continual}, the tasks contain fewer samples, as the segments are concatenated into longer utterances of 30s.
\end{enumerate}
The number of utterances per task and the task order is given for each experiment in Table \ref{tab:utterances}. 

\begin{table}
\centering
\caption{Number of utterances per task for each experiment, listed in the order in which the tasks are presented to the model.}
\setlength{\tabcolsep}{3pt}
\begin{threeparttable}
\begin{tabular}{c lr lr lr lr}
\toprule
\multicolumn{1}{c}{\textbf{Task}} &
\multicolumn{2}{c}{\textbf{Exp.~1}} &
\multicolumn{2}{c}{\textbf{Exp.~2}} &
\multicolumn{2}{c}{\textbf{Exp.~3}} &
\multicolumn{2}{c}{\textbf{Exp.~4}}\\
\midrule
1 & US  & 349{,}561 & LIB & 103{,}997 & US  & 349{,}561 & CV/E & \multicolumn{1}{c}{---} \\
2 & ENG & 116{,}464 & GB/M & 9{,}001 & RU  & 82{,}862 & CV/N & \multicolumn{1}{c}{---} \\
3 & AUS & 56{,}287 & US/U & 27{,}536 & NL  & 59{,}146 & CG/N & 27{,}363\\
4 & IND & 71{,}086 & IN/U & 9{,}271 & PL  & 92{,}221 & CG/V & 20{,}664\\
5 & SCO & 10{,}744 & IN/M & 9{,}271 & SV  & 23{,}411 & \multicolumn{2}{c}{---} \\
\midrule
\textbf{Total} & \multicolumn{2}{c}{\textbf{604{,}142}} & 
\multicolumn{2}{c}{\textbf{159{,}076}} &
\multicolumn{2}{c}{\textbf{607{,}201}} &
\multicolumn{2}{c}{\textbf{56{,}659}}\\
\bottomrule
\end{tabular}
\end{threeparttable}
\label{tab:utterances}
\end{table}

\noindent \textbf{Model.} Taking MFCC features as input, our ASR model for Experiments 1-3 consists of 12 Conformer \cite{conformers} encoder and 6 Transformer \cite{transformer} decoder layers with dimension 2048 and 4 attention heads with dimension 256. The weight of CTC is $c=0.3$ during training and decoding. For the monolingual experiments, \textit{SentencePiece} \cite{sentencepiece} is used to generate $C=5000$ output tokens on the first task; while for the multilingual experiment, it generates $C=5000$ output tokens for each task separately. In the multilingual experiment, therefore, the token embedding in the decoder and the output layers of CTC and decoder are task-specific (thus language-specific) and not shared across tasks. The remaining 92$\%$ of the parameters are shared and thus susceptible to forgetting. An overview of the number of (shared) parameters in the model and the portion affected by our method is provided in Table~\ref{tab:model}. As shown, in the monolingual experiments (Exp. 1-2), the entire model is shared and contains 46.7M parameters, of which 90.7$\%$ are in the weight matrices of the linear layers. By applying SVD to the adaptation in these matrices, and making only $\bm{\alpha}$ trainable, the number of trainable parameters is reduced to 44.2k, or less than 0.1$\%$ of total parameters. For the multilingual experiment (Exp. 3), only 43.8k parameters are trained during Stage 2 of our method, while 43.0M parameters are shared across tasks. For Experiment 4, we use the pretrained OWSM v3.2 small \cite{owsm}, which consists of eight E-Branchformer \cite{e_branchformer} encoder and eight Transformer decoder layers, and an output vocabulary of $C = 50{,}000$ tokens, amounting to 366.7M parameters. The model was pretrained on 180k hours of speech covering 151 languages, with output tokens shared between languages. When adapting OWSM to CGN, we freeze the output layers and decoder token embeddings and only update the linear weight matrices, analogous to LoRA but without the low rank component. When naively fine-tuning the model, restricting updates to linear weights led to both better learning and less forgetting, due to less overfitting. Adaptation is thus applied to 244.8M parameters (66.8$\%$ of the entire model), corresponding, as shown in Table~\ref{tab:model}, to 145.9k parameters during Stage~2 of our method. CTC is only used during training, with $c=0.3$.

\begin{table}
\centering
\caption{Number of parameters (\textit{Params.}) and percentage of total model size (\textit{Perc.}) for the shared model (shared across tasks), all (trainable) $\bm{W}$ matrices, and their corresponding $\bm{\alpha}$'s.}
\setlength{\tabcolsep}{2.5pt} % tighten inter-column spacing
\begin{threeparttable}
\begin{tabular}{l l r l r l r }
\toprule
\multirow{2}{*}{\textbf{Group}} 
 & \multicolumn{2}{c}{\textbf{Exp.~1--2}} 
 & \multicolumn{2}{c}{\textbf{Exp.~3}} 
 & \multicolumn{2}{c}{\textbf{Exp.~4}} \\
\cmidrule(lr){2-3} \cmidrule(lr){4-5} \cmidrule(lr){6-7}
 & \textbf{Params.} & \multicolumn{1}{c}{\textbf{Perc.}} & \textbf{Params.} & \multicolumn{1}{c}{\textbf{Perc.}} & \textbf{Params.} & \multicolumn{1}{c}{\textbf{Perc.}} \\
\midrule
Shared model        & 46.8M  & 100.0$\%$ & 43.0M  & 92.0$\%$ & 366.7M & 100.0$\%$ \\
All $\bm{W}$ matrices & 42.4M  & 90.7$\%$  & 39.9M  & 85.3$\%$ & 244.8M  & 66.8$\%$ \\
All $\bm{\alpha}$'s  & 44.2k  & $<$0.1$\%$ & 43.8k  & $<$0.1$\%$ & 145.9k  & $<$0.1$\%$ \\
\bottomrule
\end{tabular}
\end{threeparttable}
\label{tab:model}
\end{table}

\noindent \textbf{Training.} For Experiments 1-3, the model is trained for 80 epochs on the initial task $1$, and for 10 epochs on each subsequent tasks $2, ..., T$. In the multilingual experiment, before training the entire model on a new task (i.e. language), a new task-specific part is initialized and trained for 10 epochs. For Stage 2 training of our method, the number of epochs is set to $e_\text{st}=3$. The optimizer is Adam \cite{adam}, reset before learning a new task and before each stage. Following \cite{eeckt2021continual}, the learning rate is set 10 times smaller ($0.0001$) for tasks $2,...,T$ compared to task $1$ ($0.001$), since training does not start from scratch but from a reasonably good starting point (i.e. the previous task model parameters). During Stage 2 of our method, as $\bm{\alpha}$ is learned from scratch, we set the learning rate to $0.01$. This value ensures that, when regularization is disabled, SVR can effectively recover performance on the new task, thus allowing $\sigma(\bm{\alpha})$ to converge. For the language-specific components of the multilingual experiment, the learning rate is also set to $0.01$, which worked well to train the language-specific components of RU with the initial US model. The batch size is set to 64 at all times. For Experiment 4, OWSM is fine-tuned for 20 epochs using the Adam optimizer with a learning rate of 5.0e-5 when adapting to a new task, and a learning rate of 1.0e-2 during Stage~2 training of our method. The batch size is 64.

\noindent\textbf{Baselines}. We compare our method to state-of-the-art \text{rehearsal-based} and \text{regularization-based} CL methods for ASR, as well as to the initial model and to two methods setting the worst-case and best-case performance (though violating the principle of CL). Methods marked with \textdagger\ are rehearsal-based, requiring access to a memory of past data, as our method. 
\begin{enumerate}
    \item \textit{Initial model}. \text{Initial model} is the model prior to adaptation, i.e. with parameters $\bm{\theta}^1$. For the multilingual experiment, the task-specific parameters are trained for each task before computing Initial model's performance. The shared part of the model, however, remains fixed and only trained on the first task.
    \item \textit{Fine-Tuning}. Fine-Tuning naively adapts the model to the subsequent tasks, without taking any measures to prevent forgetting. It represents the worst-case scenario.
    \item \textit{Seperate Model (Sep. Model)}. Sep. Model keeps a separate model for each task, avoiding task interference and forgetting. The model kept per task is the model finetuned on that task with Fine-Tuning. Sep. Model thus performances as well as Fine-Tuning on new tasks but with zero forgetting (as it can still use the old models for old tasks). Although this violates the principle of CL, it represents a best-case scenario. 
    \item \textit{Experience Replay (ER)} \textdagger \cite{er}. ER trains jointly on the current mini-batch from the new task and on a mini-batch sampled from the memory, by computing the cross-entropy loss from Eq. \ref{eq:loss} for both.
    \item \textit{Knowledge Distillation (KD)} \textdagger \cite{eeckt2021continual}. Similar to ER, KD trains jointly on the new task's data and the memory $\M$. The difference with ER is that the KD loss from Eq. \ref{eq:loss_kd} is used on the mini-batch from the memory, instead of the cross-entropy loss.
    \item \textit{Learning Without Forgetting (LWF)} \cite{lwf}. LWF uses the KD loss from Eq. \ref{eq:loss_kd} on the new task's data to transfer knowledge from the old to the current model. LWF thus does not require access to a memory.
    \item \textit{Fine-Tuning with Averaging (FTA)} \cite{weight_averaging}. If $\bm{\tilde{\theta}}_t$ are the previous model parameters $\bm{\theta}_{t-1}$ fine-tuned to task $t$, then FTA computes and stores for task $t$ the model $\bm{\theta}_t = (1-\eta)\bm{\theta}_{t-1} + \eta \bm{\tilde{\theta}}_t$. Models before and after fine-tuning on task $t$ are thus averaged with a weight $\eta$. In \cite{weight_averaging}, $\eta=0.50$ or $\eta=1/t$ are tested, with the latter obtaining the best results; we therefore consider the latter.
    \item \textit{Update Only Encoders (UOE)} \cite{updating_only}. UOE suggests updating only the encoder, excluding the layer normalization layers, while freezing the rest of the model. 
    \item \textit{CL by Random Layer-wise Tuning (CLRL-Tuning)} \cite{wang23d_interspeech}. CLRL-Tuning proposes to only update $K$ randomly selected encoder layers each epoch. We set $K=1$ (the best setting in \cite{wang23d_interspeech}) and increase the number of epochs to train a new task from 10 to 50 for CLRL-Tuning. 
\end{enumerate}

\noindent \textbf{Hyper-parameters.} ER, KD, and LWF require a hyper-parameter $\lambda$ that controls the weight of the regularization loss. For LWF, we adopt the optimal value $\lambda = 0.1$ from \cite{eeckt2021continual}. For ER and KD, we select the best value between $\lambda = 0.1$ (optimal value from \cite{eeckt2021continual}) and $\lambda = 1.0$ (optimal value from \cite{vandereeckt_interspeech2023}, where the regularization loss on previous tasks is given equal weight to the cross-entropy loss on the current task). The optimal value is determined on the first adaptation of each experiment using the corresponding validation sets. The value of $\lambda$ is selected independently for each memory size $|\mathcal{M}|$.

\noindent \textbf{Metrics.} We report WER (in $\%$) per task evaluated on the model trained on all tasks. In addition, we also report Average WER, obtained by averaging the WERs of the tasks, to asses the overall performance on the model trained with the given method, and the Backward Transfer (BWT) to asses its forgetting. If $R_{i,j}$ is the WER on task $j$ of the model trained on the first $i$ tasks, then the Average WER is given by $\text{Average WER} = \frac{1}{T}\sum_{k=1}^{T} R_{T,k}$, while BWT is given by Eq. \ref{eq:bwt}.  Note that negative BWT indicates forgetting, i.e. the average increase of WER of old tasks when learning new tasks. Positive BWT, on the other hand, indicates that the model exploits the learning of new tasks to further improve its performance on old tasks. 
\begin{equation}
    \text{BWT} = \frac{1}{T-1}\sum_{k=1}^{T-1} (R_{k,k} - R_{T,k})
    \label{eq:bwt}
\end{equation}

\noindent \textbf{Significance testing}. Average WER is the main metric to compare the performance of the CL methods. Therefore,  we perform significance testing on Average WER, using Wilcoxon signed-rank test on the number of errors per utterance \cite{Strik2000ComparingTR}, considering significance levels $\alpha=0.05$ (*), $\alpha=0.01$ (**) and $\alpha=0.001$ (***), or $ns$ for non-significant ($\alpha > 0.05$).

\section{Results}
\begin{table*}
    \centering
    \begin{threeparttable}
    \caption{Results of the monolingual Exp. 1 and Exp. 2. Tasks are learned from left to right, with WERs obtained after learning all tasks. For rehearsal-based methods, the memory size $|\M|$ is indicated. SVR is our proposed method. Best Average WER with small memory ($|\M|<200$) is in bold, second best is underlined. Significance testing is applied to compare SVR against baselines and to assess the effect of memory size reduction for rehearsal-based methods (SVR, ER, KD).}
    \begin{tabular}{l r c@{\hspace{5pt}} c@{\hspace{5pt}} c@{\hspace{5pt}} c@{\hspace{5pt}} c@{\hspace{5pt}} c@{\hspace{5pt}} c@{\hspace{5.6pt}}  c@{\hspace{5pt}} c@{\hspace{5pt}} c@{\hspace{5pt}} c@{\hspace{5pt}} c@{\hspace{5pt}} c@{\hspace{5pt}} c@{\hspace{5.6pt}}}
    \toprule
    & & \multicolumn{7}{c}{\textbf{\textit{Exp. 1}}} & \multicolumn{7}{c}{\textbf{\textit{Exp. 2}}} \\
    \cmidrule(lr){3-9} \cmidrule(lr){10-16}
    \multirow{2}{*}{\textbf{Method}} & \multirow{2}{*}{$\bm{|\M|}$} & \multicolumn{5}{c}{\textbf{WER$\downarrow$ per task}} & \multicolumn{2}{c}{\textbf{Average}} & \multicolumn{5}{c}{\textbf{WER$\downarrow$ per task}} & \multicolumn{2}{c}{\textbf{Average}}   \\
    \cmidrule(lr){3-7} \cmidrule(lr){8-9}  \cmidrule(lr){10-14} \cmidrule(lr){15-16}
      & & \textbf{1--US} & \textbf{2--ENG} & \textbf{3--AUS} & \textbf{4--IND} & \textbf{5--SCO} & \textbf{WER}$\downarrow$  & \multicolumn{1}{c}{\textbf{BWT}$\uparrow$}  & \textbf{1--LIB} & \textbf{2--GB/M} & \textbf{3--US/U} & \textbf{4--IN/U} & \textbf{5--IN/M} & \textbf{WER}$\downarrow$  & \textbf{BWT}$\uparrow$ \\
    \toprule
      \multicolumn{2}{l}{Initial model} & 15.4 & 12.7 & 13.4 & 21.4 & 13.5 & 15.25 & --- & \phantom{1}6.4 & 19.4 & \phantom{1}7.9 & 18.8 & 32.9 & 17.08 & --- \\
    \multicolumn{2}{l}{Fine-Tuning} & 18.2 & 12.1 & 12.4 & 22.2 & 10.5 & 15.07 & -3.6 & 17.0 & 19.6 & 16.5 & \phantom{1}4.1 & \phantom{3}4.9 & 12.43 & -9.0 \\
    \multicolumn{2}{l}{Sep. Model} & 15.4 & 10.2 & \phantom{1}8.9 & 16.0 & 10.5 & 12.17 & \phantom{1}0.0 & \phantom{1}6.4 & \phantom{1}4.7 & \phantom{1}5.9 & \phantom{1}4.3 & \phantom{1}4.9 & \phantom{1}5.23 & \phantom{-}0.0 \\
    \midrule
    \multicolumn{2}{l}{FTA} & 15.9 & 10.9 & 10.7 & 19.3 & 11.8 & 13.71 & -0.3 & \phantom{1}7.5 & \phantom{1}8.6 & \phantom{1}8.1 & \phantom{1}8.5 & 12.2 & \phantom{1}8.97 & -0.1 \\
    \multicolumn{2}{l}{LWF} & 17.8 & 11.7 & 12.4 & 21.0 & 10.1 & 14.60 & -3.2 & 17.2 & 18.6 & 16.6 & \phantom{1}4.1 & \phantom{1}4.9 & 12.27 & -8.7 \\
    \multicolumn{2}{l}{UOE} & 18.5 & 12.5 & 12.8 & 22.6 & 10.5 & 15.36 & -3.8 & 14.3 & 19.4 & 17.1 & \phantom{1}4.4 & \phantom{1}5.4 & 12.10 & -8.2 \\
    \multicolumn{2}{l}{CLRL-Tuning} & 17.9 & 12.2 & 12.4 & 22.3 & 11.6 & 15.26 & -2.8 & 11.7 & 16.4 & 14.4 & \phantom{1}5.6 & \phantom{1}7.1 & 11.02 & -5.5 \\
    \cmidrule(lr){1-16}
    \multirow{3}{*}{KD} & $200$ & 17.9 & 11.8 & 12.1 & 20.0 & 10.4 & 14.45\tnote{d} & -2.8 & \phantom{1}7.8 & \phantom{1}6.9 & \phantom{1}8.0 & \phantom{1}4.6 & \phantom{1}5.6 & \phantom{1}6.60\tnote{d} & -1.1 \\
     & 20 & 18.2 & 12.1 & 12.2 & 21.0 & 10.6 & 14.81\tnote{d,f} & -3.3 & \phantom{1}9.5 & \phantom{1}9.9 & \phantom{1}9.9 & \phantom{1}4.5 & \phantom{1}5.3 & \phantom{1}7.82\tnote{d} & -2.8  \\
     & $1t$ & 18.3 & 12.3 & 12.2 & 20.8 & 20.4 & 14.81\tnote{f} & -3.2 & 10.8 & 12.6 & 11.1 & \phantom{1}4.7 & \phantom{1}5.6 & \phantom{1}8.92\tnote{d} & -4.1 \\
    \cmidrule(lr){1-16}
     \multirow{3}{*}{ER} & $200$ & 17.9 & 12.1 & 12.0 & 20.2 & 10.5 & 14.54\tnote{f} & -2.9 & \phantom{1}7.6 & \phantom{1}6.8 & \phantom{1}7.9 & \phantom{1}4.5 & \phantom{1}5.4 & \phantom{1}6.43\tnote{d} & -1.1 \\
      & 20 & 17.8 & 11.8 & 12.3 & 20.8 & 10.5  & 14.62\tnote{e,f} & -3.0 & \phantom{1}9.4 & 10.4 & \phantom{1}9.9 & \phantom{1}4.3 & \phantom{1}5.1 & \phantom{1}7.82\tnote{d} & -3.0 \\
      & $1t$ & 18.2 & 11.8 & 12.5 & 21.0 & 10.5 & 14.80\tnote{e} & -3.2 & 12.2 & 13.7 & 11.8 & \phantom{1}4.2 & \phantom{1}5.1 & \phantom{1}9.39\tnote{d} & -5.1 \\
    \midrule
     \multirow{2}{*}{SVR} & $20$ & 15.9 & 10.4 & 10.4 & 18.3 & 10.9 & \textbf{13.18}\tnote{a,f} & -0.6 & \phantom{1}8.3 & \phantom{1}7.9 & \phantom{1}8.8 & \phantom{1}5.8 & \phantom{1}7.7 & \phantom{1}\underline{7.70}\tnote{b,f} & -1.1 \\
      & $1t$ & 16.0 & 10.5 & 10.5 & 18.6 & 11.0 & \underline{13.33}\tnote{a,f} & -0.7 & \phantom{1}8.4 & \phantom{1}8.2 & \phantom{1}8.7 & \phantom{1}5.5 & \phantom{1}7.3 & \phantom{1}\textbf{7.62}\tnote{c,f}  & -1.3 \\
    \bottomrule
    \end{tabular}
    \begin{tablenotes}
    \footnotesize
    \item[a] Significantly outperforms all baselines, including those with large memory $|\M|=200$, with level ***.
    \item[b] Significantly outperforms all baselines with small ($|\M| < 200$) or no memory with level ***, except KD with $|\M|=20$ (\textit{ns}).
    \item[c] Significantly outperforms all baselines with small ($|\M| < 200$) or no memory with level ***, except KD with $|\M|=20$ with level *.
    \item[d, e] Reducing memory significantly deteriorates performance of given method, with level *** for \textit{d} and * for \textit{e}.
    \item[f] Reducing memory does not significantly deteriorate performance of given method (level is \textit{ns}).
    \end{tablenotes}
    \label{tab:full_results}
    \end{threeparttable}
\end{table*}

\begin{table*}
\centering
\begin{threeparttable}
\caption{Results of multilingual Exp.~3 and OWSM-based Exp.~4. Tasks are learned from left to right, with WERs obtained after learning all tasks. For rehearsal-based methods, the memory size $|\M|$ is indicated. SVR is our proposed method. Best Average WER with small memory ($|\M|<200$) is in bold, second best is underlined. Significance testing is applied to compare SVR against baselines and to assess the effect of memory size reduction for rehearsal-based methods (SVR, ER, KD).}
\begin{tabular}{l r
c@{\hspace{7pt}} c@{\hspace{7pt}} c@{\hspace{7pt}} c@{\hspace{7pt}} c@{\hspace{7pt}} c@{\hspace{7pt}} c@{\hspace{7pt}}
c@{\hspace{7pt}} c@{\hspace{7pt}} c@{\hspace{7pt}} c@{\hspace{7pt}} c@{\hspace{7pt}} c@{\hspace{7pt}}}
\toprule
& &
\multicolumn{7}{c}{\textbf{\textit{Exp. 3}}} &
\multicolumn{6}{c}{\textbf{\textit{Exp. 4}}} \\
\cmidrule(lr){3-9} \cmidrule(lr){10-15}
\multirow{2}{*}{\textbf{Method}}
& \multirow{2}{*}{$\bm{|\M|}$}
& \multicolumn{5}{c}{\textbf{WER$\downarrow$ per task}} & \multicolumn{2}{c}{\textbf{Avg.}}
& \multicolumn{4}{c}{\textbf{WER$\downarrow$ per task}} & \multicolumn{2}{c}{\textbf{Avg.}} \\
\cmidrule(lr){3-7} \cmidrule(lr){8-9}
\cmidrule(lr){10-13} \cmidrule(lr){14-15}
& &
\textbf{1--US} & \phantom{U}\textbf{2--RU} & \textbf{3--NL} & \textbf{4--PL} & \textbf{5--SV} &
\textbf{WER$\downarrow$} & \textbf{BWT$\uparrow$} &
\textbf{1--CV/E} & \textbf{2--CV/N} & \textbf{3--CG/N} & \textbf{4--CG/V} &
\textbf{WER$\downarrow$} & \textbf{BWT$\uparrow$} \\
\toprule
\multicolumn{2}{l}{Initial model} 
& 15.4 & 45.7 & 27.6 & 29.8 & 44.8 & 32.64 & --- 
& 13.4 & 15.4 & 45.7 & 37.9 & 28.08 & --- \\
\multicolumn{2}{l}{Fine-Tuning} 
& 44.6 & 47.1 & 18.0 & 14.6 & 22.1 & 29.29 & -14.2 
& 24.3 & \phantom{1}9.6 & 27.7 & 13.6 & 18.81 & -2.4 \\
\multicolumn{2}{l}{Sep. Model} 
& 15.4 & 26.3 & 13.6 & 12.2 & 22.1 & 17.90 & \phantom{-1}0.0
& 13.4 & 15.4 & 22.4 & 13.6 & 16.18 & \phantom{-}0.0 \\
\midrule
\multicolumn{2}{l}{FTA} 
& 19.9 & 30.8 & 18.0 & 18.5 & 29.0 & 23.22 & \phantom{1}-0.7
& 14.8 & \phantom{1}9.5 & 26.8 & 18.5 & 17.42 & +2.2 \\
\multicolumn{2}{l}{LWF} 
& 50.7 & 62.3 & 19.3 & 14.0 & 20.8 & 33.39 & -19.9
& --- & --- & --- & --- & --- & --- \\
\multicolumn{2}{l}{CLRL-Tuning} 
& 24.8 & 43.5 & 23.3 & 21.5 & 31.8 & 28.99 & \phantom{1}-4.1
& --- & --- & --- & --- & --- & --- \\
\cmidrule(lr){1-15}
\multirow{2}{*}{KD}
& $200$ 
& 27.2 & 32.2 & 16.1 & 14.2 & 23.4 & 22.61\tnote{e} & \phantom{1}-5.3
& 18.7 & \phantom{1}9.7 & 26.0 & 13.5 & 16.98\tnote{e} & -0.3 \\
& $20$ 
& 44.3 & 39.8 & 17.2 & 14.3 & 22.2 & 27.55\tnote{e} & -12.0
& 20.4 & 10.5 & 26.0 & 13.6 & 17.64\tnote{e} & -1.6 \\
\cmidrule(lr){1-15}
\multirow{2}{*}{ER}
& $200$ 
& 25.9 & 34.0 & 17.5 & 14.6 & 23.2 & 23.04\tnote{e} & \phantom{1}-5.6
& 17.4 & \phantom{1}8.9 & 27.0 & 13.4 & 16.72\tnote{e} & -0.2 \\
& $20$ 
& 33.9 & 39.6 & 18.2 & 15.6 & 22.8 & 26.00\tnote{e} & \phantom{1}-9.7
& 19.5 & 10.6 & 26.3 & 13.3 & 17.41\tnote{e} & -1.6 \\
\midrule
\multirow{3}{*}{SVR}
& $50$ 
& 21.2 & 31.1 & 17.1 & 16.4 & 27.2 & \textbf{22.60}\tnote{c,f} & \phantom{1}-1.2
& 18.0 & \phantom{1}8.4 & 24.4 & 14.9 & \textbf{16.42}\tnote{d,g} & +0.4 \\
& $20$ 
& 21.9 & 31.5 & 17.1 & 16.5 & 27.6 & \underline{22.93}\tnote{b,f,h} & \phantom{1}-1.8
& 18.7 & \phantom{1}8.5 & 24.5 & 15.0 & \underline{16.65}\tnote{a,g,e} & +0.2  \\
& $1t$ 
& 23.5 & 31.5 & 17.2 & 16.2 & 27.5 & 23.15\tnote{b,h} & \phantom{1}-1.7
& 18.9 & \phantom{1}8.8 & 25.9 & 14.8 & 17.10\tnote{a,e} & -1.9 \\
\bottomrule
\end{tabular}
\begin{tablenotes}
\footnotesize
    \item[a, b] Significantly outperforms all baselines with small ($|\M| < 200$) or no memory with level *** (\textit{a}), except FTA with level \textit{ns} (\textit{b}).
    \item[c] Significantly outperforms all baselines with level ***, except FTA with level ** and KD with $|\M|=200$ with level \textit{ns}.
    \item[d] Significantly outperforms all baselines with level ***, except ER with $|M|=200$ with level $ns$.
    \item[e, f, g] Reducing memory significantly deteriorates performance of given method with level *** for \textit{e}, level ** for \textit{f} and level * for \textit{g}.
    \item[h] Reducing memory does not significantly deteriorate performance of given method (level is \textit{ns}).
\end{tablenotes}
\label{tab:exp3_exp4}
\end{threeparttable}
\end{table*}

\begin{figure}
    \centering
    \subfigure[Exp. 1]{
        \includegraphics[width=0.90\linewidth]{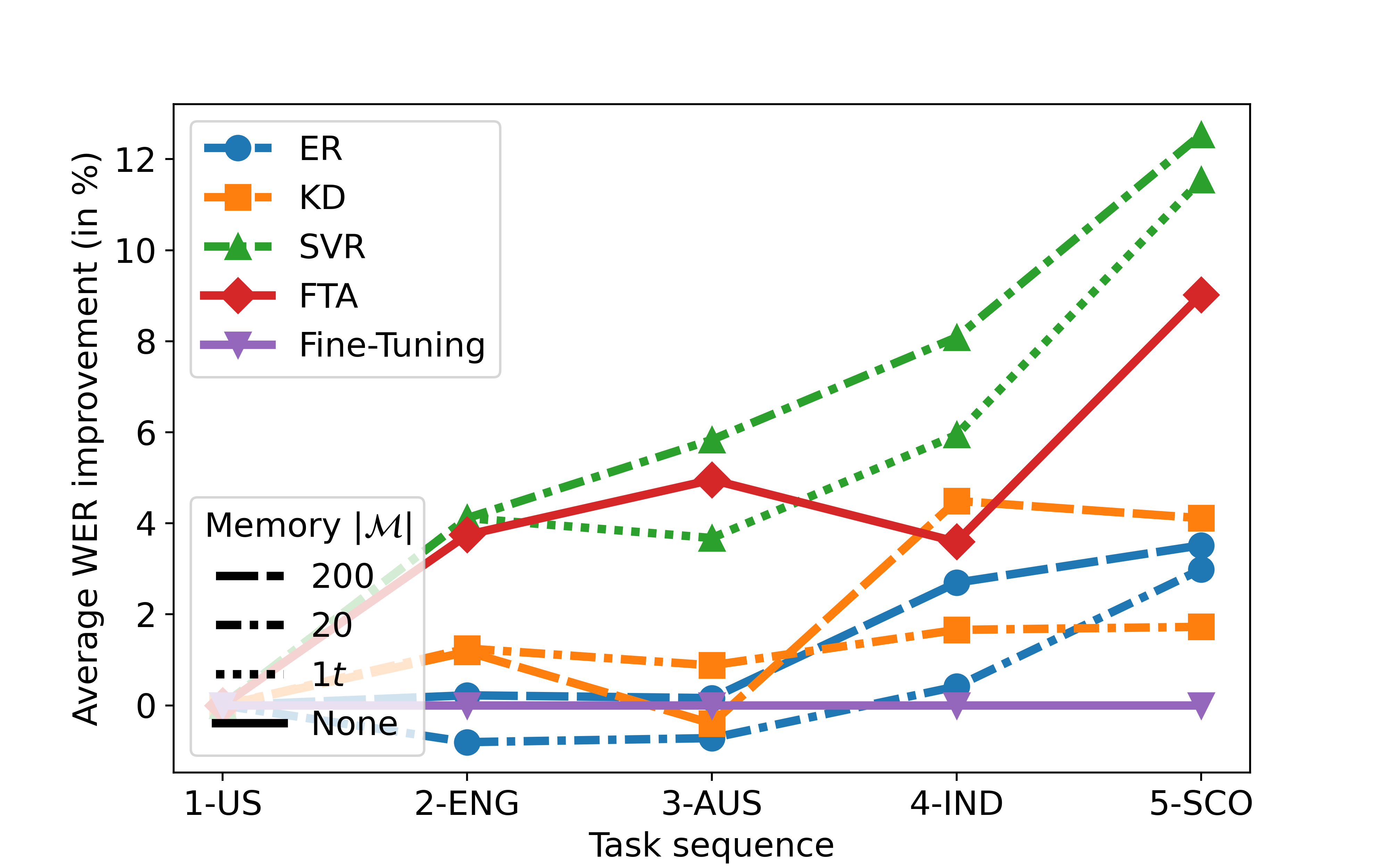}
        \label{fig:exp1_avgwer}
    }
    \hfill
    \subfigure[Exp. 2]{
        \includegraphics[width=0.90\linewidth]{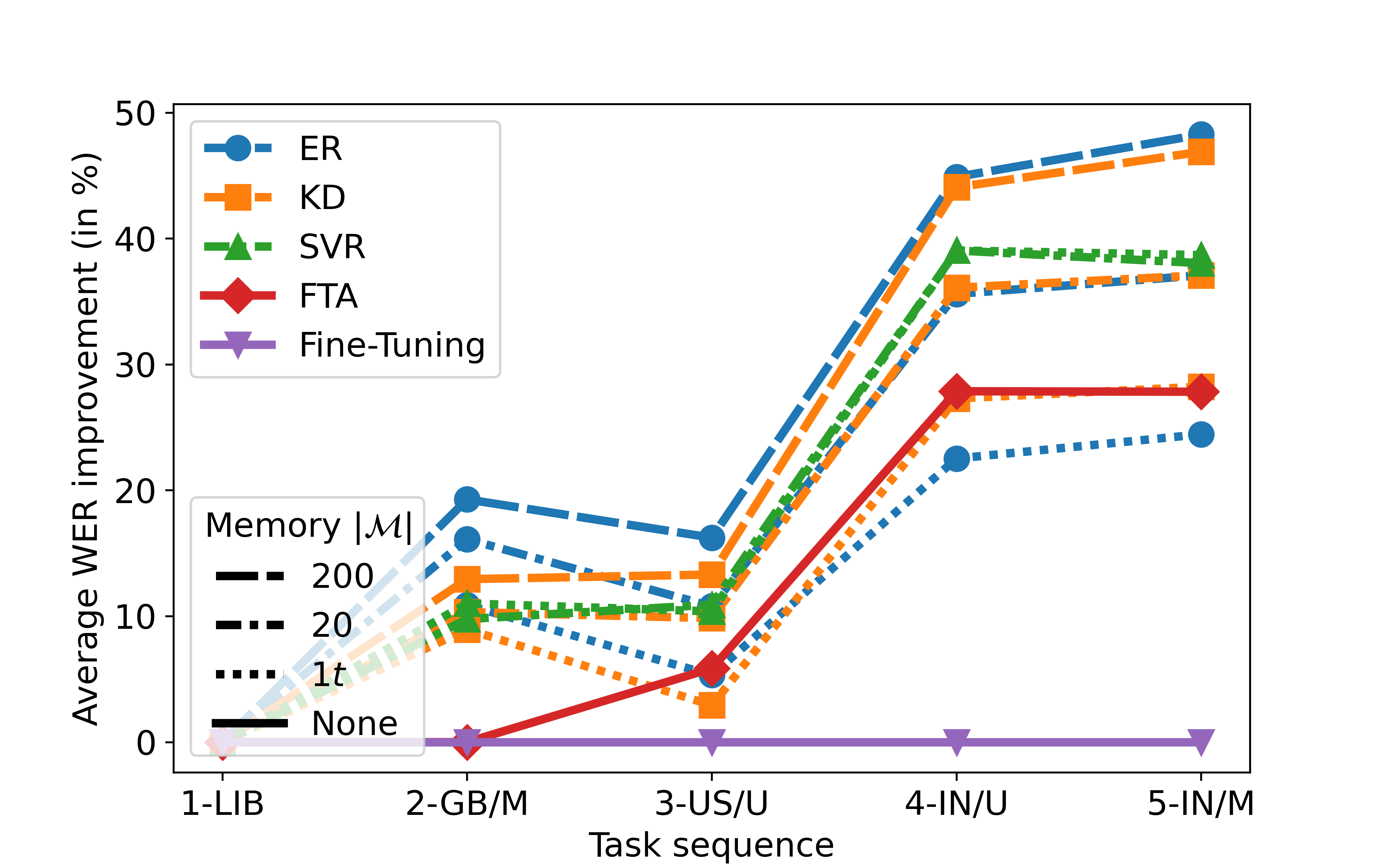}
        \label{fig:exp2_avgwer}
    }
    \hfill
    \subfigure[Exp. 3]{
        \includegraphics[width=0.90\linewidth]{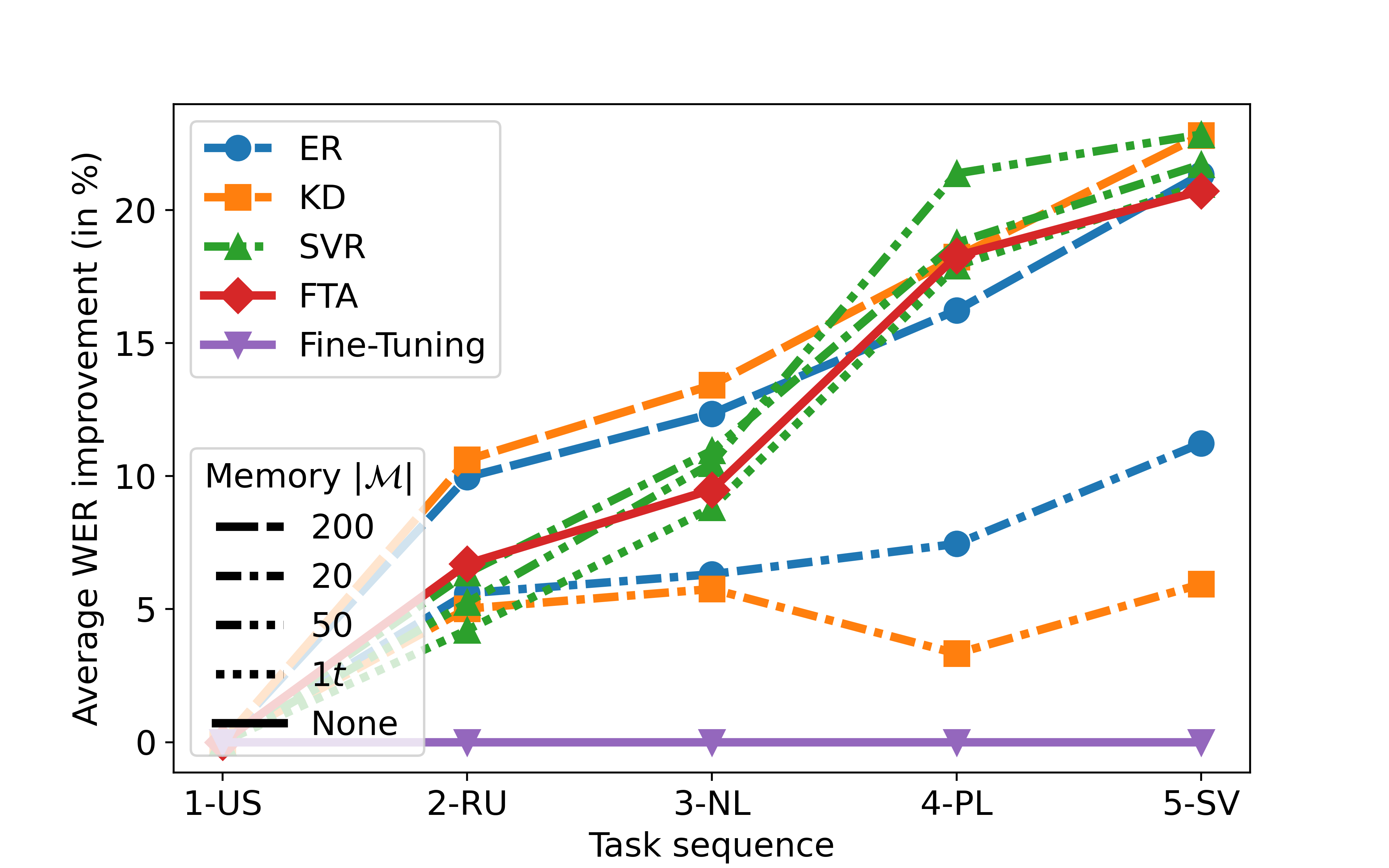}
        \label{fig:exp3_avgwer}
    }
    \caption{Average WER improvement (\%) over Fine-Tuning, reported per method and after each of the five tasks. The values indicate the relative percentage reduction in Average WER compared to Fine-Tuning at each task.}
    \label{fig:avg_wer}
\end{figure}

\subsection{Experiment 1}
\label{subsec:exp1}
The results for Exp.~1 are shown in Table~\ref{tab:full_results} and visualized in Figure~\ref{fig:exp1_avgwer}. We evaluate SVR with memory sizes $|\M|=20$ and $|\M|=1t$, and compare against baselines ER and KD, each tested with memory sizes $|\M|=200$, $|\M|=20$ and $|\M|=1t$. We draw the following observations.

Regarding the performance of baselines, we observe that, due to the diversity in speakers, accents, and lexical content in Common Voice, even rehearsal-based methods such as ER and KD perform relatively poorly. With a memory of $|\M|=200$ utterances, \textit{KD and ER reduce forgetting by only 1/5th}, and neither approach comes close to the best case performance given by \textit{Sep. Model}. When the memory is reduced to $|\M|=20$ utterances, ER and KD only marginally improve over Fine-Tuning, offering limited mitigation of forgetting. On the other hand, the \textit{non-rehearsal-based baselines} such as LWF, CLRL-Tuning, and UOE fail to substantially improve upon Fine-Tuning. While LWF yields slight improvements, forgetting remains substantial, and the performance gap to \textit{Sep. Model} persists. The exception is FTA, which nearly eliminates forgetting, but at the cost of limited learning on new tasks—especially as the number of tasks $t$ increases, since the weight of new tasks decreases proportionally ($1/t$). 

\textit{Our method, SVR, outperforms FTA, as well as ER and KD—even with a large memory of $|\M|=200$ utterances}, in addition achieving better final performance on four out of five tasks. This holds not only with, for SVR, access to a small memory of $|\M|=20$ utterances, but even with access to just \textit{one utterance per task} ($|\M|=1t$). In both cases, \textit{SVR reduces forgetting compared to Fine-Tuning by more than $80\%$}, while still achieving high performance on new tasks. Unlike FTA, \textit{SVR strikes a much better balance between retaining old knowledge and integrating new information}. Overall, SVR closes the gap between Fine-Tuning (worst case) and Sep. Model (base case) by nearly \textit{two-thirds} using $|\M|=20$, and by over \textit{60\%} using $|\M|=1t$. 

Figure~\ref{fig:exp1_avgwer} shows the relative improvement in Average WER over Fine-Tuning after each task. \textit{SVR with 20 memory samples consistently outperforms all other methods after each stage}. While FTA performs competitively in the early stages, it fails to adapt well as more tasks are introduced, and the gap between FTA and SVR widens over time. Notably, \textit{SVR with just one memory sample per task outperforms KD and ER with $|\M|=200$ utterances after all tasks}, despite the $200\times$ (when learning 2--ENG) to $50\times$ (when leaning 5--SCO) difference in memory size.

\subsection{Experiment 2}
\label{subsec:exp2}
The results for Exp. 2 are reported in Table \ref{tab:full_results} and visualized in Figure~\ref{fig:exp2_avgwer}. While the overall trends are consistent with those observed in Exp. 1 (Sec. \ref{subsec:exp1}), some important distinctions arise due to the nature of the tasks involved.

Compared to Common Voice, the LibriSpeech and Libri-Adapt tasks consist of much longer utterances with less variation in terms of speakers and lexical content. For example, comparing 2--ENG from Exp. 1 and 2--GB/M from Exp. 2, we find that the former includes 1.8k speakers, 58.4k unique words, and 9.2 words per utterance, while the latter contains 211 speakers, 19.8k unique words, and 34.88 words per utterance. This difference is reflected in the results: in Exp. 2, both ER and KD perform strongly with $|\M|=200$, and even with $|\M|=20$, they still yield substantial improvements over Fine-Tuning. Fewer memory samples are thus required to maintain strong performance.

Despite this, SVR remains highly competitive. \textit{With a memory of just $|\M|=20$, SVR continues to outperform both ER and KD, and even with only one utterance per task ($|\M|=1t$), it exceeds their performance when they use 20 samples}. In this low-memory regime, ER and KD suffer significant degradation, whereas SVR maintains robust performance—forgetting approximately 60$\%$ less than ER and KD when the latter use a larger memory of $|\M|=20$ utterances. Relative to Fine-Tuning, SVR reduces forgetting by more than $85\%$ with just one utterance per task.

SVR also outperforms all non-rehearsal-based methods. As in Exp. 1, LWF, UOE, and CLRL-Tuning offer only modest gains over Fine-Tuning, and fail to match the effectiveness of memory-based approaches. FTA again nearly eliminates forgetting, but does so at the cost of poor plasticity—its ability to learn new information is significantly constrained.

Figure~\ref{fig:exp2_avgwer} illustrates the relative improvement in Average WER over Fine-Tuning after each task. From the third task onward (3--US/U), SVR with $|\M|=20$ and $|\M|=1t$ outperform both KD and ER with $|\M|=20$. This reflects a general trend: while SVR forgets less, it may also learn new tasks slightly less than ER and KD. This trade-off is visible in the first adaptation (to 2--GB/M), where ER and KD enjoy an initial advantage in Average WER due to their higher plasticity. However, the gap narrows and eventually reverses as more tasks are learned—particularly because SVR better preserves prior knowledge, a factor that becomes increasingly important in later stages of continual learning.

\subsection{Experiment 3}
\label{subsec:exp3}
Table~\ref{tab:exp3_exp4} presents the results of the multilingual experiment (Exp.~3). Several insights can be drawn from these results.

First, the \textit{Initial model}, which combines the initial shared model trained on US-English with adapted language-specific components for each target language, performs poorly on all languages except US. This highlights the need to adapt the shared parameters, which account for approximately 92$\%$ of the total model parameters (see Table~\ref{tab:model}), in order to achieve strong multilingual performance.

Next, \textit{SVR demonstrates a clear advantage} over existing rehearsal-based methods, even with minimal memory. With $|\M|=1t$, SVR substantially outperforms both ER and KD configured with $|\M|=20$, reducing their forgetting by \textit{between 82$\%$ and 85$\%$}. SVR also significantly outperforms non-rehearsal baselines such as LWF and CLRL-Tuning.

Compared to rehearsal-based methods with larger memory ($|\M|=200$), SVR with $|\M|=20$ or $|\M|=1t$ still achieves a \textit{reduction in forgetting of 65$\%$ to 70$\%$}. However, in terms of average WER, it merely matches ER and slightly underperforms KD, which benefits from greater flexibility and hence learns new tasks more effectively. Nonetheless, we expect SVR’s reduced forgetting to yield larger benefits as more tasks are added. This trend is already visible in Figure~\ref{fig:exp3_avgwer}, which plots average WER improvement over Fine-Tuning after each task. After tasks 4 and 5, the performance gap between SVR ($|\M|=20$ and $|\M|=1t$) and the large-memory baselines (ER, KD with $|\M|=200$) has noticeably narrowed, suggesting that \textit{SVR’s advantage may increase with longer task sequences}.

Despite these strengths, SVR with small memory ($|\M|=20$ or $|\M|=1t$) does not significantly outperform FTA. As previously discussed in Sections~\ref{subsec:exp1} and~\ref{subsec:exp2}, FTA achieves strong performance through \textit{minimal forgetting}, but at the cost of limited learning capacity due to the fixed averaging weight of $1/t$. In this multilingual experiment, however, the impact of FTA’s rigidity appears mitigated by the presence of \textit{3.8M task-specific parameters} (Table~\ref{tab:model}), which help each new language adapt without substantial updates to the shared model. Consequently, while SVR with $|\M|=20$ and $|\M|=1t$ achieve average WERs that are \textit{1.2$\%$ and 0.3$\%$ better than FTA}, respectively, these differences are not statistically significant.

Increasing SVR’s memory size to $|\M|=50$ yields a notable improvement in this multilingual setting. The average WER is reduced by an additional \textit{1.4$\%$} compared to $|\M|=20$, reflecting both improved task learning—particularly on the final task (SV)—and \textit{33$\%$ less forgetting}. As shown in Figure~\ref{fig:exp3_avgwer}, this improvement only becomes apparent from task 4 onward. At earlier stages, when the memory holds 10–20 utterances from prior tasks (e.g., RU and US), SVR with $|\M|=20$ performs comparably to $|\M|=50$. However, at task 4, when only 6–7 utterances per previous task remain in memory, and at task 5, with just 5 per task, the limited rehearsal data appears insufficient to maintain SVR’s effectiveness.

Crucially, SVR with $|\M|=50$ now \textit{outperforms all baselines significantly, including FTA and ER with $|\M|=200$}, except KD with $|\M|=200$, which it \textit{matches in performance despite requiring four times less memory}.

\subsection{Experiment 4}
Table \ref{tab:exp3_exp4} presents the results for the OWSM-based experiment (Exp. 4). We make the following observations.

First, although initial task 2--CV/N was part of the OWSM v3.2 small pre--training corpus, its performance improves when learning task 3--CG/N, even under fine-tuning. This indicates \textit{positive backward transfer} \cite{gem} from 3--CG/N to 2--CV/N. Since the degradation on 2--CV/N when adapting to 4--CG/V is smaller than this gain for all tested methods, the final performance on 2--CV/N remains higher than initially, thus increasing BWT. For FTA, and SVR with $|\mathcal{M}|=50$ and $|\mathcal{M}|=20$, the positive backward transfer on 2--CV/N outweighs cumulative forgetting, yielding a positive BWT value. After four tasks, SVR exploits this effect most strongly, reducing the WER on 2--CV/N to 8.4\% (with $|\M|=50$), \textit{showing that SVR has the greatest ability to leverage positive backward transfer}.

For FTA the BWT is also highly positive, mainly because forgetting on 1--CV/E is limited; however, as in the other experiments, FTA struggles to learn new tasks. SVR forgets more on 1--CV/E than FTA (which could be alleviated by increasing the importance of CV/E in the memory), but overall \textit{finds a better balance between retaining prior knowledge and learning new tasks}, achieving a lower Average WER and outperforming FTA on all Dutch tasks for every memory size.

Compared to ER and KD, SVR forgets substantially less and maintains a better stability–plasticity balance on the three Dutch tasks. With $|\mathcal{M}|=50$, SVR outperforms KD with $|\mathcal{M}|=200$ and matches ER with $|\mathcal{M}|=200$, while SVR with $|\mathcal{M}|=20$ or $|\mathcal{M}|=1t$ outperforms KD and ER with their smaller $|\mathcal{M}|=20$ memory. SVR therefore \textit{achieves competitive or superior performance while relying on four--to--five times fewer stored utterances}, and its performance degrades more gracefully than ER and KD
as the memory decreases.

\subsection{Impact of Memory size on Performance}

\begin{figure}
    \centering
    \subfigure[Experiment 1. The $x$-axis is logarithmically scaled.]{
        \includegraphics[width=0.90\linewidth]{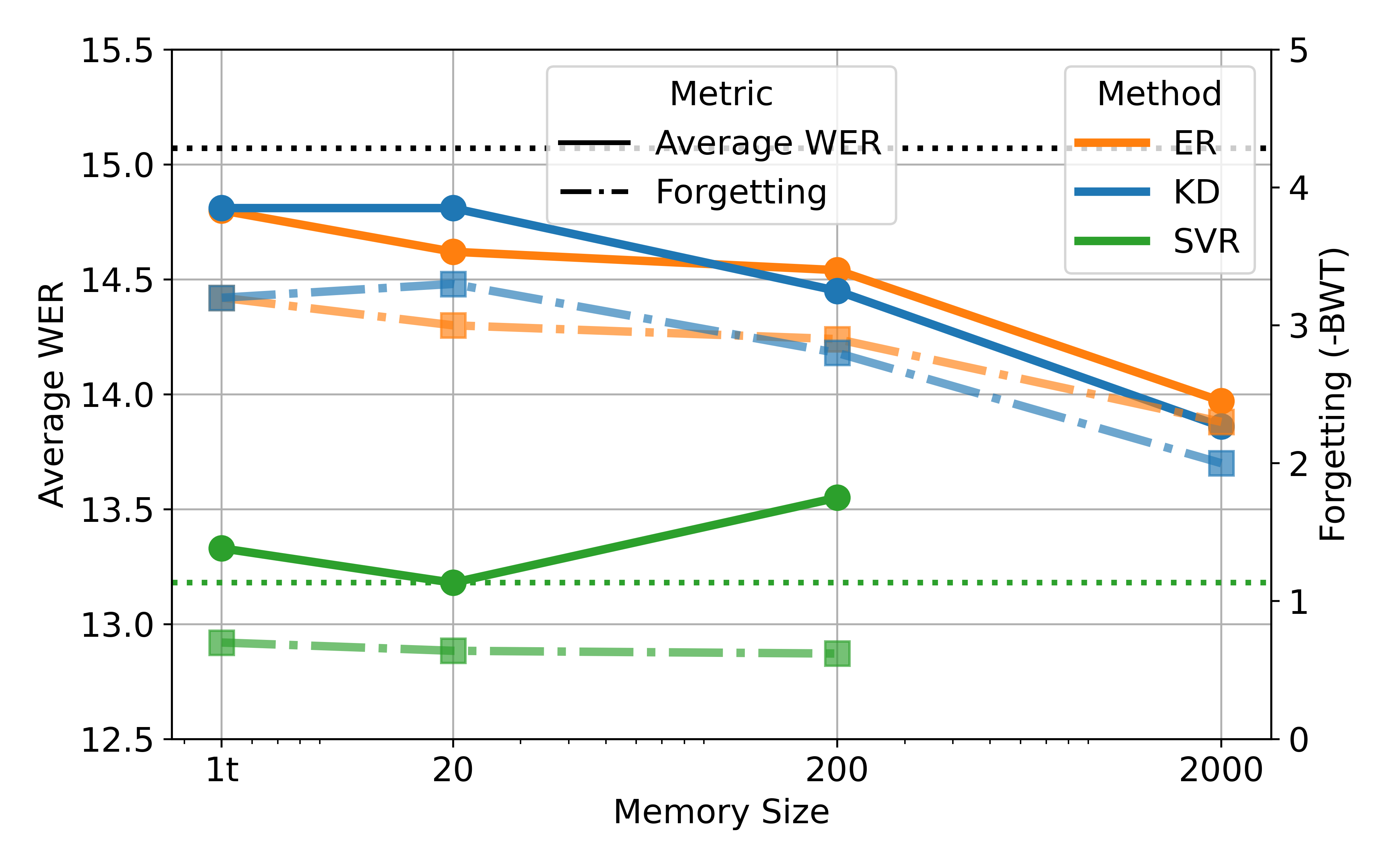}
        \label{fig:memory_sizes:exp1}
    }
    \hfill
    \subfigure[Experiment 3]{
        \includegraphics[width=0.90\linewidth]{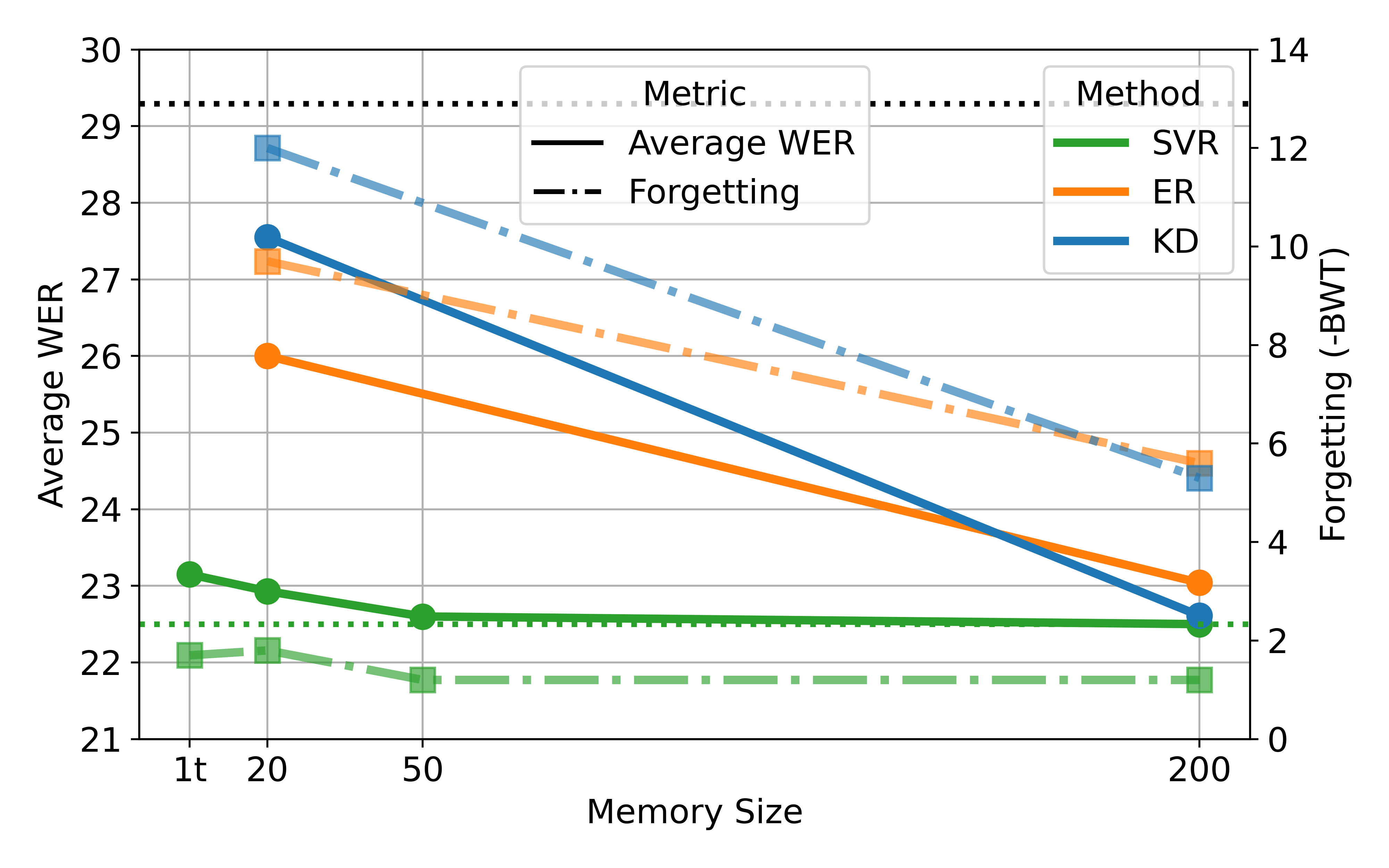}
        \label{fig:memory_sizes:exp3}
    }
    \caption{Comparison of performance of rehearsal-based baselines (KD and ER) and our method (SVR) for different memory sizes $|\M|$. On the left, the y-axis shows the Average WER, where lower is better. On the right, it shows Forgetting, equal to -BWT, such that, here too, lower is better.}
    \label{fig:memory_sizes}
\end{figure}

Tables~\ref{tab:full_results} and ~\ref{tab:exp3_exp4} report results for KD, ER, and our method SVR under varying memory sizes. These results are further summarized in Figure~\ref{fig:memory_sizes:exp1} for Exp. 1 and Figure \ref{fig:memory_sizes:exp3} for the multilingual Exp. 3. In both plots, the average WER  (left) and Forgetting (right, equal to -BWT) are shown as a function of the memory size for SVR, ER, and KD. Several trends are apparent from these comparisons.

For Exp. 1 (Figure \ref{fig:memory_sizes:exp1}, Table~\ref{tab:full_results}), we observe that the performance of ER and KD improves consistently as the memory size increases from $|\M|=1t$ to $|\M|=2000$. At small memory sizes ($|\M|=1t$ and $|\M|=20$), their performance is only marginally better than Fine-Tuning, but the gap widens with larger memory. However, even at $|\M|=2000$, both KD and ER still \textit{underperform SVR with much smaller memory sizes}—namely, $|\M|=20$ and even $|\M|=1t$; moreover, \textit{KD's and ER's forgetting is, for all memory sizes, much higher than SVR's.}  For ER, each memory increase leads to a statistically significant improvement in performance, except from $|\M|=20$ to $|\M|=200$. For KD, too, increasing the memory size generally results in significant improvement, except from $|\M|=1t$ to $|\M|=20$, which both only marginally improve Fine-Tuning's performance. For SVR, increasing the memory size from $|\M|=1t$ to $|\M|=20$ does not result in a significant performance difference. Moreover, increasing the memory size from $|\M|=20$ to $|\M|=200$ even results in deterioration in terms of average WER, which can be attributed to reduced flexibility due to a more challenging regularization loss. Up until 3--AUS, SVR with $|\M|=200$ outperforms SVR with $|\M|=20$ with an average WER of 11.65 vs. 11.78 and a positive BWT of +0.1 vs. -0.2, i.e. less forgetting. However, from 4--IND onward, SVR with $|\M|=200$ starts underperforming SVR with $|\M|=20$. Although its forgetting is still less (BWT of -0.1 vs. -0.3), it learns the new task not as well (18.3 WER on IND vs. 17.5) due to a $16\%$ lower mean $\sigma(\bm{\alpha})$ (averaged over all linear layers) of 0.32 vs. 0.38 for SVR with $|\M|=20$. When learning task 5--SCO, the mean $\sigma(\bm \alpha$) is 0.20 for SVR with $|\M|=200$ compared to 0.28 for SVR with $|\M|=20$. Reducing the weight of the regularization loss (from Eq. \ref{eq:loss_st2}) for SVR with $|\M|=200$ would result in larger mean $\sigma(\bm \alpha)$ and facilitate its learning of new tasks, thus improving its performance. However, while the regularization weight from Eq.~\ref{eq:loss_st2} appears suboptimal for larger memory sizes, it is important to note that our method is designed for and tailored to scenarios with extremely limited memory. In conclusion, we find that our method is highly effective even under minimal memory constraints, and forgets significantly less than ER and KD—even when these baselines use memory sizes as large as $|\M|=2000$.

For Exp. 3 (Figure \ref{fig:memory_sizes:exp3}, Table~\ref{tab:exp3_exp4}), similar trends emerge. Increasing the memory from $|\M|=20$ to $|\M|=200$ results in a significant improvement for both ER and KD. Nevertheless, \textit{SVR with a memory of $|\M|=50$ outperforms these baselines} -- w.r.t. ER even a memory of $|\M|=20$ suffices for SVR, despite their use of four to ten times more memory. Moreover, \textit{SVR forgets between $68\%$ and $78\%$ less} (even with the smallest memory), and thus, as more tasks would be added and forgetting becomes more crucial, SVR's performance is expected to improve further in relation to ER and KD. For SVR, the increase from $|\M|=1t$ to $|\M|=20$ is not statistically significant, but the subsequent step to $|\M|=50$ yields a significant performance gain. As discussed in Sec.~\ref{subsec:exp3} and shown in Figure~\ref{fig:exp3_avgwer}, this improvement becomes apparent only from task 4 onward. For the earlier tasks, the smaller memory ($|\M|=20$) remains sufficient to sustain SVR’s effectiveness. Increasing $|\M|$ from $|\M|=50$ to $|\M|=200$ does not yield significant improvement.  

In Exp. 2 (Table \ref{tab:full_results}), memory reductions negatively impact performance for ER and KD. Specifically, reducing the memory from $|\M|=200$ to $|\M|=20$, and further to $|\M|=1t$, leads to consistent and significant degradation. In contrast, \textit{SVR maintains stable performance when reducing memory from $|\M|=20$ to $|\M|=1t$}, and even in the smallest-memory setting ($|\M|=1t$), it significantly outperforms KD and ER configured with $|\M|=20$.

Finally, in Exp.~4 (Table~\ref{tab:exp3_exp4}), the performance of SVR remains stable when reducing the memory size, although the degradation is significant. In the larger OWSM-based model with multilingual setting, SVR clearly benefits from a larger memory, yet—unlike ER and KD—its performance does not collapse when the memory is reduced.

In summary, across all three experiments, SVR exhibits strong robustness to reduced memory size and consistently delivers superior or comparable performance to competing methods, while forgetting much less; even when operating under substantially stricter memory constraints.

\subsection{Analysis of $\sigma(\bm{\alpha})$}

\begin{figure}
    \centering
    \subfigure[Mean $\sigma(\bm{\alpha})$ per layer]{
        \includegraphics[width=1.00\linewidth]{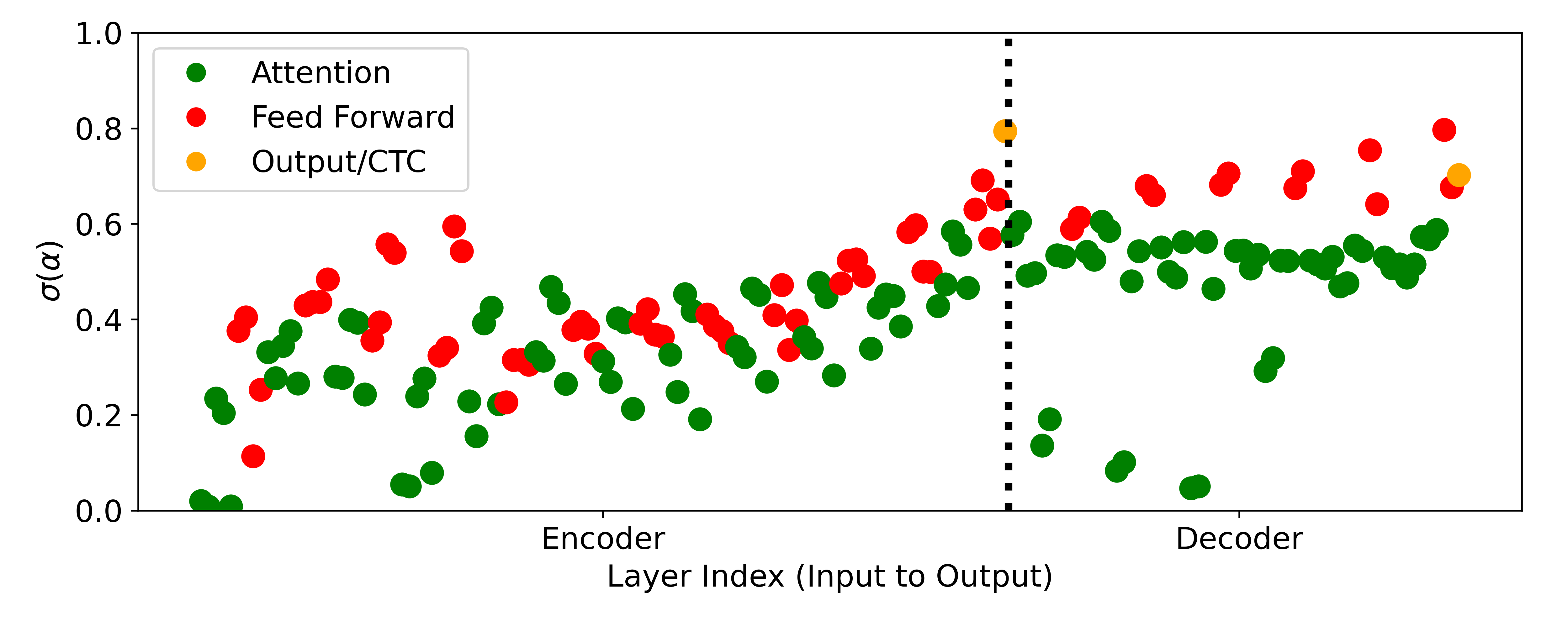}
        \label{fig:alphas_mean}
    }
    \hfill
    \subfigure[Percentage of $\sigma(\bm{\alpha})$ for which $0.05 <\sigma(\alpha_i)< 0.95$]{
        \includegraphics[width=1.00\linewidth]{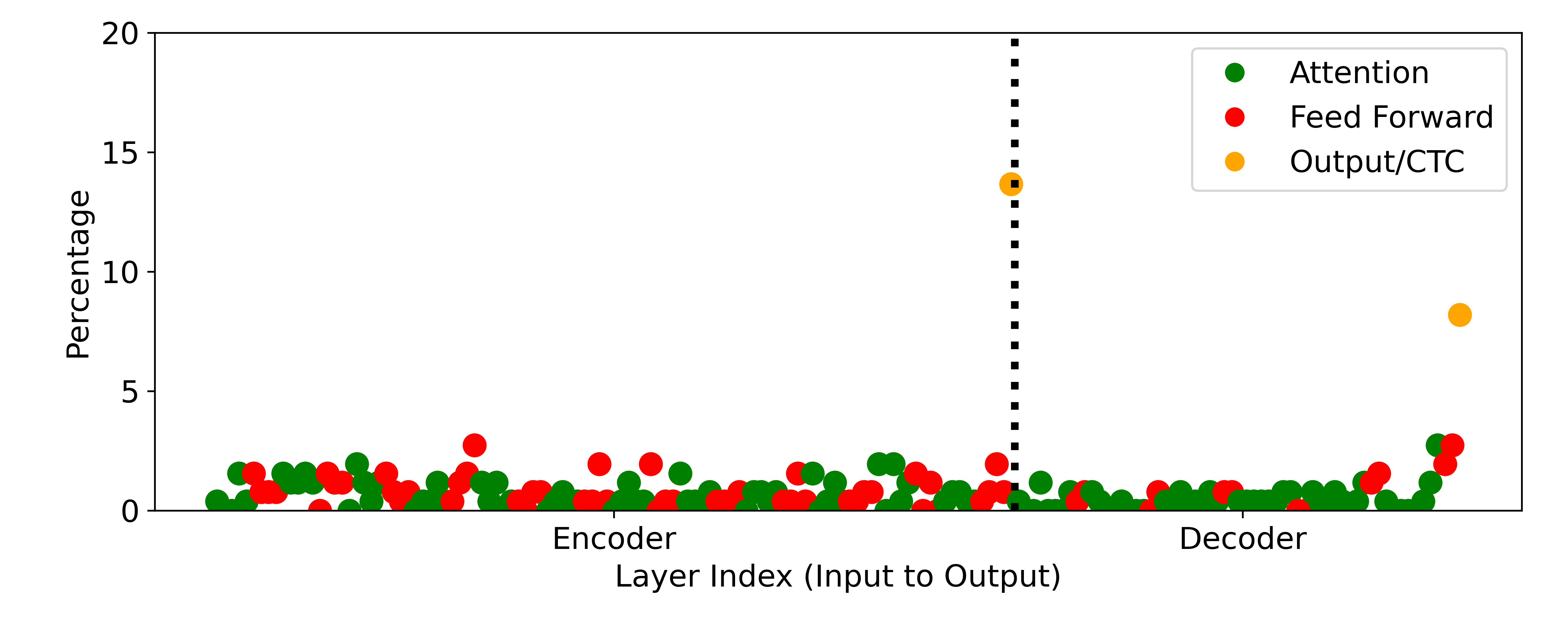}
        \label{fig:alphas_ratio}
    }
    \caption{Plot of $\sigma(\bm{\alpha})$ across layers (to which our method is applied) for SVR [$|\M|=1$] after the first adaptation (1--US $\rightarrow$ 2--ENG) of Exp. 1. The layers are sorted from input (left) to output (right), with the dotted line indicating the end of encoder and start of decoder. Layers are colored based on layer type. Fig. \ref{fig:alphas_mean} gives the mean $\sigma(\bm{\alpha})$ value per layer, while Fig. \ref{fig:alphas_ratio} gives the percentage of $\sigma(\bm{\alpha})$ which differ significantly (by more than 0.05) from 0 or 1, i.e. the percentage of $\sigma(\bm{\alpha})$ for which $0.05 < \sigma(\alpha_i)<0.95$. }
    \label{fig:alphas_heatmap}
\end{figure}

\begin{figure}
    \centering
    \subfigure[Mean $\sigma(\bm{\alpha})$ per layer]{
        \includegraphics[width=0.95\linewidth]{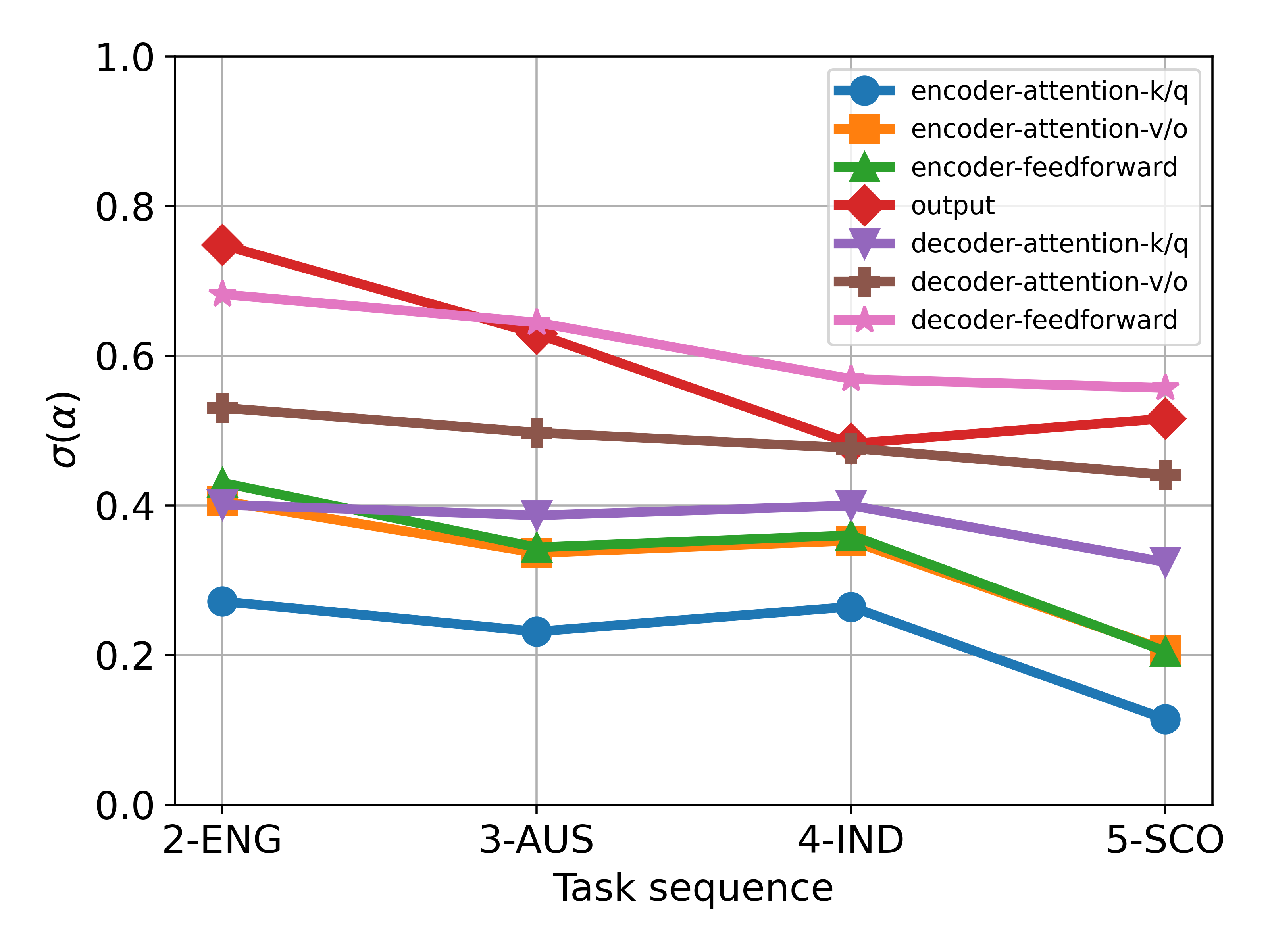}
        \label{fig:alphas_mean_over_time}
    }
    \hfill
    \subfigure[Percentage of $\sigma(\bm{\alpha})$ for which $0.05 <\sigma(\alpha_i)< 0.95$]{
        \includegraphics[width=0.95\linewidth]{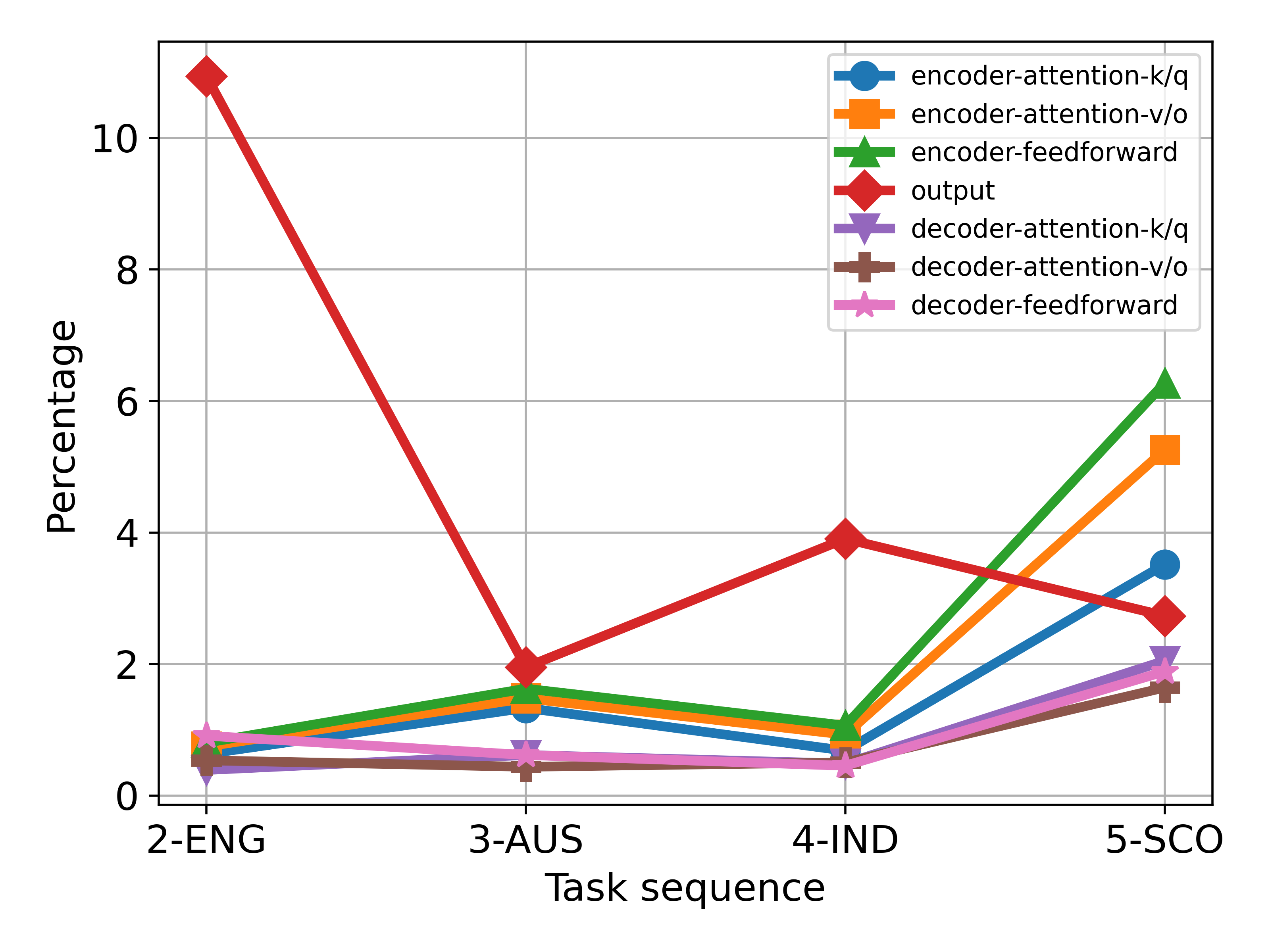}
        \label{fig:alphas_ratio_over_time}
    }
    \hfill
    \subfigure[Percentage of $\sigma(\bm{\alpha})$ for which $\sigma(\alpha_i)< 0.05$]{
        \includegraphics[width=0.95\linewidth]{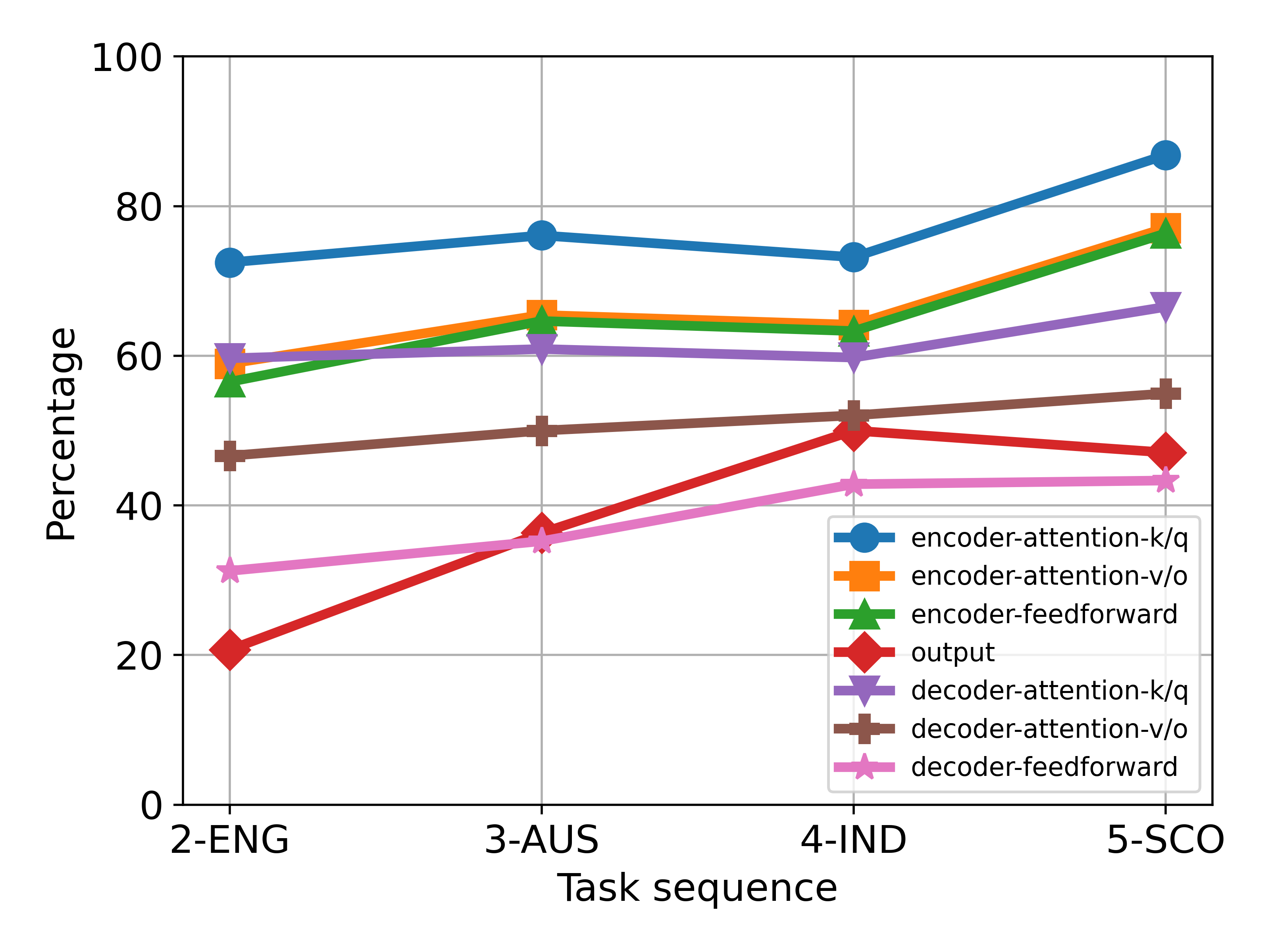}
        \label{fig:alphas_zeroratio_over_time}
    }
    \caption{Mean and percentage of intermediate and small values of $\sigma(\bm{\alpha})$ across layer groups after learning each task for SVR [$|\M|{=}1$] on Exp. 1. The layer groups are: (1) encoder key-query (k/q) attention;  (2) encoder value-output (v/o) attention; (3) encoder feedforward; (4) output layers (CTC and decoder output); (5) decoder key-query (k/q) attention;  (6) decoder value-output (v/o) attention; and (7) decoder feedforward.
Fig. ~\ref{fig:alphas_mean_over_time} shows the mean $\sigma(\bm{\alpha})$ per group after learning each task.
Fig. ~\ref{fig:alphas_ratio_over_time} shows the percentage of parameters in $\sigma(\bm{\alpha})$ for which $0.05 < \sigma(\alpha_i) < 0.95$.
Fig. ~\ref{fig:alphas_zeroratio_over_time} shows the percentage for which $\sigma(\alpha_i) < 0.05$.}
    \label{fig:alphas_plot_over_time}
\end{figure}

To gain deeper insight into the behavior of our method, we analyze the learned values of $\sigma(\bm{\alpha})$ across layers and tasks. Since $\sigma(\bm{\alpha})$ controls the extent to which each singular value-based rank-one update is applied, examining its distribution helps illuminate how the model balances retaining old knowledge and integrating new information. This analysis not only enhances understanding of our method’s internal mechanisms but also reveals which parts of the model are more affected by inter-task conflicts and adaptation needs.

Figure~\ref{fig:alphas_heatmap} presents plots of $\sigma(\bm{\alpha})$ after the first adaptation step in Exp. 1 (i.e., from task 1--US to task 2--ENG), using SVR with $|\M|=1t$. Specifically, Figure~\ref{fig:alphas_mean} shows the mean value of $\sigma(\bm{\alpha})$ for each layer to which our method is applied, while Figure~\ref{fig:alphas_ratio} displays the percentage of parameters for which $0.05 < \sigma(\alpha_i) < 0.95$, indicating the extent of intermediate, non-binary gating behavior. The layers are ordered from input (left) to output (right), with the dotted line marking the boundary between the encoder and decoder components.

From Figure~\ref{fig:alphas_mean}, we observe that the CTC output layer exhibits the highest mean $\sigma(\bm{\alpha})$ values, approaching 0.80. This suggests that rank-one updates in this layer are largely retained during adaptation, contributing significantly to learning the new task without severely impacting performance on the old task. A clear trend is also visible across the network: deeper encoder and decoder layers generally show higher $\sigma(\bm{\alpha})$ values, while earlier layers tend to suppress more updates. Within the decoder, feedforward submodules exhibit higher $\sigma(\bm{\alpha})$ values than their attention-related counterparts, indicating that the feedforward layers play a more prominent role in accommodating the new task.

Turning to Figure~\ref{fig:alphas_ratio}, we find that most $\sigma(\bm{\alpha})$ values are concentrated near 0 or 1, with relatively few falling in the intermediate range. This binary-like behavior suggests a strong preference for either fully applying or completely discarding rank-one updates. The proportion of intermediate values ($0.05 < \sigma(\alpha_i) < 0.95$) is generally low, with output layers showing the highest ratio, around 13$\%$ maximum. For all other layers, this proportion remains below 3$\%$. These findings contrast with weight averaging \cite{weight_averaging}, which uniformly scales updates (e.g., by 0.5 or $1/t$); in our method, the model instead learns to make more decisive gating choices, favoring either full integration or full rejection of the learned updates.

Figure~\ref{fig:alphas_plot_over_time} extends this analysis by showing how $\sigma(\bm{\alpha})$ evolves across tasks. In this figure, layers are grouped into seven functional categories: (1) encoder key-query attention;  (2) encoder value-output attention; (3) encoder feedforward; (4) output layers (CTC and decoder output); (5) decoder key-query attention;  (6) decoder value-output attention; (7) decoder feedforward layers. We split the attention block into key-query and value-output, hypothesizing that the value and output weight matrices in the attention function analogously to feedforward layers, and therefore expect their $\sigma(\bm \alpha)$ behavior to resemble that of the feedforward layers rather than that of the query and key weight matrices.

Figure~\ref{fig:alphas_mean_over_time} shows the mean value of $\sigma(\bm{\alpha})$ per group after each task in Exp. 1. The highest mean values are consistently found in the output, decoder feedforward and decoder value-output attention layers, suggesting that updates in these layers are most beneficial for adapting to new tasks while preserving performance on past ones. Conversely, the encoder shows the lowest mean values, indicating either that updates in these layers conflict more with previously learned knowledge or that they are less important for learning new tasks. Across tasks, the mean $\sigma(\bm{\alpha})$ values decrease slightly, but remain relatively high compared to weight averaging, which would hyperbolically decrease adaptation weights as $1/t$.

Figure~\ref{fig:alphas_ratio_over_time} tracks the ratio of $\sigma(\bm{\alpha})$ values in the intermediate range, that is, values of $\sigma(\bm{\alpha})$ for which $0.05<\sigma(\alpha_i)<0.95$. These remain consistently low across all layer groups and tasks, in most cases below 4$\%$, except for the output layers after the first adaptation ($\approx11\%$), and the encoder feedforward and value-output attention layers at the last adaptation ($\approx 6\%$). This supports the earlier observation that the model tends to make sharp gating decisions rather than compromise between preserving and discarding updates.

Finally, Figure~\ref{fig:alphas_zeroratio_over_time} illustrates the proportion of $\sigma(\bm{\alpha})$ values below 0.05, effectively indicating how many updates are nearly fully suppressed. This ratio increases slightly over time for all layer groups. Notably, encoder layers and decoder key-value attention layers show suppression rates above 50$\%$, reinforcing the notion that these layers are more susceptible to inter-task interference or less useful for new-task adaptation. In contrast, the output, decoder value-output attention and decoder feedforward layers initially suppress fewer updates (well below 40$\%$), though this ratio gradually increases as more tasks are learned. 

In summary, we observe that the model typically exhibits a binary behavior in its use of rank-one updates—either fully suppressing or fully accepting them. Suppression is most pronounced in the encoder and decoder key-query attention layers, while acceptance is highest in the output layers, decoder value-output attention and decoder feedforward layers. This selective behavior appears to stem from two factors. First, high suppression rates may be observed in layers where task interference is substantial—here, rank-one updates that could benefit the new task are nonetheless suppressed due to their detrimental impact on performance for previous tasks. Second, suppression may also occur in more task-agnostic layers, where adaptation offers only marginal gains for the new task while risking (slight) degradation for the old ones. In both cases, the model attempts to find a balance between plasticity (adaptation to the new task) and stability (not forgetting of old tasks) by filtering out updates that harm old tasks more than they improve the performance on the new task.

\subsection{Ablation Study}
\label{subsec:ablation}
\begin{table}
    \centering
    \begin{threeparttable}
    \caption{Ablation study of our method. Rows marked with ``$+$'' indicate cumulative changes to the configuration listed directly above, while rows marked with ``$\rightarrow$'' represent alternative variants of the reference method.}
    \begin{tabular}{l c@{\hspace{5pt}} c@{\hspace{5pt}}}
    \toprule
    \multirow{2}{*}{\textbf{Model}} & \multicolumn{2}{c}{\textbf{Average}} \\
    \cmidrule(lr){2-3}
     & \textbf{WER}$\downarrow$ & \textbf{BWT}$\uparrow$  \\
    \toprule
    \multicolumn{3}{c}{\textit{Exp. 1. Full adaptation 1--US $\rightarrow\dots \rightarrow$ 5--SCO}} \\
    \cmidrule(lr){1-3}
    SVR [$|\M|=20$] & $13.18$ & -0.6 \\
    \phantom{x}$+$ Memory loss not scaled with $t-1$ & $13.54$\tnote{a} & -1.4 \\
    \phantom{x}$\rightarrow$ No $\L_\text{kd}$ on memory, only $\L_\text{ce}$ (Eq.~\ref{eq:loss_st2}) & $13.89$\tnote{b} & -2.0 \\
    \phantom{x}$\rightarrow$ No $\L_\text{ce}$ on memory, only $\L_\text{kd}$ (Eq.~\ref{eq:loss_st2}) & $13.83$\tnote{c} & -1.5 \\
        \cmidrule(lr){1-3}
    ER [$|\M|=200$] & 14.54 & -2.9 \\
    \phantom{x}$\rightarrow$ Use KD instead of ER & 14.45\tnote{d} & -2.8 \\
    \phantom{x}$\rightarrow$ Combine KD and ER & 14.47\tnote{d} & -2.7 \\
    \cmidrule(lr){1-3}
    SVR [$|\M|=1t$] & $13.33$ & -0.7   \\
     \phantom{x}$\rightarrow$ Rest of model not averaged, old params used & $13.85$\tnote{b} & -0.3 \\
     \phantom{x}$\rightarrow$ Rest of model not averaged, new params used & $13.84$\tnote{b} & -1.5  \\
    \midrule
    \multicolumn{3}{c}{\textit{Exp. 2. First adaptation 1--LIB $\rightarrow$ 2--GB/M}} \\
    \cmidrule(lr){1-3}
     SVR [$|\M|=1$] & \phantom{1}$6.46$ & -0.8    \\
     \phantom{x}$\rightarrow$ $\sigma(\alpha_i)$ not restricted to $[0, 1]$ & \phantom{1}$6.62$\tnote{c} & -1.8 \\
     \midrule
     \multicolumn{3}{c}{\textit{Exp. 2. Fourth adaptation 1--LIB $\rightarrow\dots\rightarrow$ 4--IN/U}} \\
    \cmidrule(lr){1-3}
    FTA: fixed averaging weight $\eta = 1/t$                 & 8.77 & -0.9 \\
    \phantom{x}$+$ Learn $\eta$ using ER with $|\M|=1t$     & 8.75\tnote{g}   & -3.9 \\
     \phantom{x}$+$ Learn $\eta$ using our loss (Eq.~\ref{eq:loss_st2}) instead of ER & 8.61\tnote{f} & -1.3 \\
     \phantom{x}$+$ Learn scalar $\sigma(\alpha)$ per linear layer instead of $\eta$ & 7.88\tnote{e} & -1.4  \\
    \phantom{x}$+$ SVR: learn $\bm{\alpha}\in\mathbb{R}^k$ per linear layer & {7.41}\tnote{e} & {-1.3} \\
    \bottomrule    
    \end{tabular}
    \begin{tablenotes}
    \footnotesize
    \item[a] Significant deterioration (level ***) w.r.t. the method above.
    \item[b, c, d] Deterioration w.r.t. the reference method is significant with level *** for \textit{b} and level * for \textit{c}, and non-significant (\textit{ns}) for \textit{d}. 
    \item[e, f, g] Improvement  w.r.t the method above is significant with level *** for \textit{e} and with level ** for \textit{f}, and non-significant (\textit{ns}) for \textit{g}.
    \end{tablenotes}
    \label{tab:ablation}
    \end{threeparttable}
\end{table}

Table~\ref{tab:ablation} presents an ablation study of our method, focusing on Exp. 1 and Exp. 2, providing the following insights.

\subsubsection{Stage 2 loss} When the weight of the regularization term in the stage 2 loss (see Eq.~\ref{eq:loss_st2}) is not scaled with $t{-}1$, \textit{forgetting increases} and overall performance deteriorates, highlighting the importance of scaling regularization with the number of old tasks. If, in addition, the KD loss is omitted and only the cross-entropy loss is used in Eq.~\ref{eq:loss_st2}, \textit{forgetting further increases} and performance degrades even more; replacing the cross-entropy loss with KD alone does not improve performance either, emphasizing the benefit of combining ER (through the cross-entropy loss) with KD in our formulation.

\subsubsection{Combining ER and KD in standard rehearsal}
Starting from ER with $|\mathcal{M}|=200$, we observe that neither replacing the cross-entropy loss with knowledge distillation (as in KD), nor combining cross-entropy and distillation losses (as in \cite{dark_er}), closes the performance gap with SVR.

\subsubsection{Remaining parameters} The treatment of the remaining 9.3$\%$  model parameters (from Exp. 1-3)—those not part of the $\bm{W}$ matrices of linear layers—also has a significant effect. If these parameters are not averaged during the parameter merging step (Lines~\ref{line:averaging_start}–\ref{line:averaging_end} in Algorithm~\ref{alg:overview}), but instead the old values are retained (i.e., replacing Line~\ref{line:averaging} with $\bm{p} \gets \bm{p}_{t-1}$), forgetting is more than halved. However, this comes at the cost of reduced plasticity, ultimately leading to worse average performance. Conversely, using the new parameters (i.e., $\bm{p} \gets \bm{\tilde{p}}_t$) increases the ability to learn the current task but \textit{more than doubles forgetting}, resulting in even worse overall performance. These observations suggest that even the remaining 9.3$\%$ of parameters significantly affect SVR's balance between learning and forgetting, and extending SVR to include other layer types could yield further improvements.

\subsubsection{Restricting gating vectors.} Removing the constraint that $\sigma(\bm{\alpha})$ lies within $[0, 1]$ gives the model more freedom to learn new tasks and improves short-term learning. However, it \textit{more than doubles forgetting}, due to overfitting on the memory. This confirms that constraining $\sigma(\bm{\alpha})$ is essential for maintaining generalization within the small memory constraint.

\subsubsection{From FTA to SVR} Compared to FTA, SVR learns a gating vector $\sigma(\bm{\alpha})$ per linear layer, corresponding to one gating scalar per rank-one parameter update, rather than fixing a single model-wise weight $\eta = 1/t$. By incrementally adding the components of SVR [$|\M| = 1t$] to FTA, we analyze the contribution of each improvement. We evaluate this on the fourth adaptation of Exp.~2, which—as indicated by the large performance gap between Fine-Tuning and the CL methods in Fig.~\ref{fig:exp2_avgwer}—is particularly challenging due to (a) the high zero-shot WER on 4--IN/U and (b) the latter's high interference with previous tasks. Results show that learning $\eta$ instead of fixing it significantly improves performance, though only when using our loss from Eq.~\ref{eq:loss_st2}. In this case, learning $\eta$ closes 11$\%$ of the gap between FTA and SVR. Replacing the model-wise averaging weight $\eta$ with a (linear) layer-wise scalar $\sigma(\alpha)$ provides additional flexibility—allowing some linear layers to accept updates while others suppress them—closing an additional 54$\%$ of the performance gap between FTA and SVR. Finally, learning a gating vector $\sigma(\bm{\alpha})$ per linear layer rather than a single scalar allows selective control over individual update directions within each layer, yielding a further 35$\%$ reduction of the gap and matching SVR’s full performance.
 
\section{Discussion}

Our results demonstrate that \textit{Singular Value-based Rehearsal (SVR) offers a strong and versatile solution for continual learning in ASR, particularly under tight memory constraints}. Across all three experiments—two monolingual and one multilingual—\textit{SVR consistently achieved the best performance} among all tested methods. While it required a slightly larger memory size in the multilingual setting to match its monolingual performance, it remained significantly more memory-efficient than the rehearsal-based baselines.

One of \textit{SVR’s core strengths lies in its ability to balance stability and plasticity}: it forgets substantially less than ER and KD while learning new tasks more effectively than methods such as FTA. his performance is enabled by a structured and interpretable adaptation mechanism: SVR learns for each rank-one parameter update whether to fully suppress or accept it, rather than partially weighting it, as is done in weight averaging \cite{weight_averaging}. An analysis of this binary behavior gives insights into where adaptation is most beneficial to the new task or most harmful to the old tasks.

Despite its strengths, there are avenues for improving SVR. 

In the multilingual setting, a larger memory was required to reach the same level of performance relative to the baselines as in the monolingual setups. This suggests that additional regularization on $\sigma(\bm{\alpha})$ values may help better constrain the model's flexibility when few examples per task are available. 

Another limitation is that SVR is currently applied only to the weight matrices of linear layers—which covers the majority of model parameters, but not all. Our ablation study (Sec.~\ref{subsec:ablation}) showed that the remaining 9.3$\%$ of parameters (from Exp. 1-3) can significantly affect forgetting and overall performance. Extending SVR to these components could further improve results, particularly in more heterogeneous or modern architectures. Alternatively, for these components a single scalar $\sigma(\alpha_\text{avg}) \in [0, 1]$ could be trained to learn the weight of the averaging automatically and replace Line \ref{line:averaging} in Algorithm \ref{alg:overview} with $\bm{p} \gets (1-\sigma(\alpha_\text{avg}))\bm{p}_{t-1}+\sigma(\alpha_\text{avg})\bm{\tilde{p}}_t$.

In addition, while SVR effectively exploits the available memory, it currently uses uniform sampling for memory population. A central challenge for any rehearsal-based continual learning method lies in the dual task of selecting and exploiting memory samples effectively. Given strict memory constraints, such methods must not only leverage the limited samples to their fullest potential during training -- in which SVR succeeds -- but also ensure that the most representative or informative examples are selected for storage. We currently rely on uniform sampling to populate the memory, without any mechanism to prioritize sample selection. Future work could therefore improve our approach by investigating smarter sampling, which may further enhance performance.

Moreover, combining our method with parameter-efficient fine-tuning (PEFT) methods such as LoRA \cite{lora, xu24h_interspeech} and other PEFT-techniques \cite{corda, wang2025ssvd} could improve both scalability and performance. Similarly, integrating complementary continual learning strategies—such as regularization- approaches—into the initial stage of our method may further reduce forgetting, especially in more adversarial task sequences.

Although our method was developed and evaluated in the context of end-to-end ASR, it is inherently model-agnostic and not specific to the speech domain. Applying it to other areas such as natural language processing or computer vision—where continual learning challenges also arise—could help demonstrate its broader utility and robustness.

\section{Conclusion}

In this work, we introduced Singular Value-based Rehearsal (SVR), a memory-efficient rehearsal method for continual learning (CL) in ASR. SVR is designed to operate effectively even under extreme memory constraints—a common limitation in real-world applications where storage, privacy, or model accessibility pose significant challenges.

By decomposing the linear weight updates from task-specific fine-tuning into their singular value components, and training a gating vector to modulate these updates based on a small rehearsal set, SVR offers a principled mechanism for balancing retention and adaptation. Our extensive experiments across monolingual and multilingual ASR settings show that SVR consistently outperforms existing rehearsal-based and regularization-based methods, achieving state-of-the-art performance even under minimal memory constraints.

Extensive ablation studies further underscore the importance of each component in SVR—ranging from the use of knowledge distillation, the averaging of non-linear layer weights, to the constrained gating on singular values. Moreover, the method's interpretability reveals a distinctive binary gating behavior, selectively retaining only updates that are beneficial for the new task without sacrificing performance on past ones.

In summary, SVR delivers a strong trade-off between plasticity and stability in continual ASR learning. It is scalable, interpretable, and operates effectively under strict memory limitations. While we have focused on ASR, the method is broadly applicable to other domains where continual learning and efficient adaptation are required. Future work may expand SVR to other parameter types, integrate smarter memory selection strategies, or explore its potential within large-scale models and non-speech modalities.

\section*{Acknowledgments}
Research supported by Research Foundation Flanders (FWO) under grant S004923N of the SBO programme.


\bibliographystyle{ieeetr}  % Change "plain" to "unsrt" or "abbrv"
\bibliography{main}

%\newpage

 
%\begin{IEEEbiography}[{\includegraphics[width=1in,height=1.25in,clip,keepaspectratio]{steven.jpeg}}]{Steven Vander Eeckt}
%Steven Vander Eeckt obtained a Master of Science in Mathematical Engineering (burgerlijk ingenieur) from the KU Leuven in 2019 and a Master of Science in Data Science, Big Data from Université libre de Bruxelles in 2020. Since September 2020, he is a PhD student at Departement of Electrical Engineering ESAT-PSI at KU Leuven. His PhD topic is Continual Learning in Speech Recognition. 
%\end{IEEEbiography}

%\begin{IEEEbiography}[{\includegraphics[width=1in,height=1.25in,clip,keepaspectratio]{hugo.png}}]{Hugo Van hamme}
%Hugo Van Hamme (M’92–SM’11) received the master’s degree in engineering (burgerlijk ingenieur) from Vrije Universiteit Brussel (VUB), Brussels, Belgium, in 1987, the M.Sc. degree from Imperial College London, London, U.K, in 1988, and the Ph.D. degree in electrical engineering from VUB, in 1992. From 1993 to 2002, he was with LH Speech Products and ScanSoft, initially as a Senior Researcher and later as a Research Manager. Since 2002, he has been a Professor with the Department of Electrical Engineering, KU Leuven, Leuven, Belgium. His main research interests include automatic speech assessment, assistive speech technology, source separation, and noise robust speech recognition, models of language acquisition, and computer-assisted learning.
%\end{IEEEbiography}

\vfill

\end{document}